\documentclass[nohyperref]{article}

\usepackage{microtype}
\usepackage{graphicx}
\usepackage{subfigure}
\usepackage{booktabs} 
\usepackage{makecell}
\usepackage{float}
\usepackage{multirow}
\usepackage{bbm}

\usepackage{hyperref}

\newcommand{\model}{MolDiff}

\usepackage[accepted]{icml2023}

\usepackage{amsmath}
\usepackage{amssymb}
\usepackage{mathtools}
\usepackage{amsthm}

\usepackage[capitalize,noabbrev]{cleveref}

\theoremstyle{plain}

\theoremstyle{definition}

\theoremstyle{remark}

\usepackage[textsize=tiny]{todonotes}

\icmltitlerunning{Addressing the Atom-Bond Inconsistency Problem in 3D Molecule Generation}

\begin{document}

\twocolumn[
\icmltitle{MolDiff: Addressing the Atom-Bond Inconsistency Problem in 3D Molecule Diffusion Generation}




\begin{icmlauthorlist}
\icmlauthor{Xingang Peng}{pku,pkuai}
\icmlauthor{Jiaqi Guan}{uiuc}
\icmlauthor{Qiang Liu}{uta}
\icmlauthor{Jianzhu Ma}{thu}
\end{icmlauthorlist}

\icmlaffiliation{pku}{School of Intelligence Science and Techology, Peking University, Beijing, China}
\icmlaffiliation{pkuai}{Institute for Artifical Intelligence, Peking University, Beijing, China}
\icmlaffiliation{uiuc}{Department of Computer Science, University of Illinois Urbana-Champaign,  Champaign, USA}
\icmlaffiliation{uta}{University of Texas at Austin, Texas, USA}
\icmlaffiliation{thu}{Institute for AI Industry Research, Tsinghua University, Beijing, China}
\icmlcorrespondingauthor{Jianzhu Ma}{majianzhu@tsinghua.edu.cn}

\icmlkeywords{3D Molecule Generation, Diffusion}

\vskip 0.3in
]



\printAffiliationsAndNotice{}  

\begin{abstract}
Deep generative models have recently achieved superior performance in 3D molecule generation. Most of them first generate atoms and then add chemical bonds based on the generated atoms in a post-processing manner. 
However, there might be no corresponding bond solution for the temporally generated atoms as their locations are generated without considering potential bonds. We define this problem as the atom-bond inconsistency problem and claim it is the main reason for current approaches to generating unrealistic 3D molecules. 
To overcome this problem, we propose a new diffusion model called \model{} which can generate atoms and bonds simultaneously while still maintaining their consistency by explicitly modeling the dependence between their relationships.
We evaluated the generation ability of our proposed model and the quality of the generated molecules using criteria related to both geometry and chemical properties. The empirical studies showed that our model outperforms previous approaches, achieving a three-fold improvement in success rate and generating molecules with significantly better quality.

\end{abstract}

\section{Introduction}

Small molecule drugs perform biological functions by binding to particular protein pockets based on their 3D structures. Recently, both academia and industry started to focus on designing small molecule drugs in the 3D space instead of only using 2D graphs \citep{gebauer2019symmetry, luo2022autoregressive, equivariant_diffusion}. For instance, deep learning models have been proposed to generate multiple 3D molecule conformations which have the potential to be further developed as drugs \cite{xu2022geodiff, guan2021energy} or can directly bind given protein pockets\cite{luo20213d, li2021structure, liu2022graphbp, peng2022pocket2mol}. 

\begin{figure}[t]
    \centering
    \includegraphics[width=0.98\linewidth]{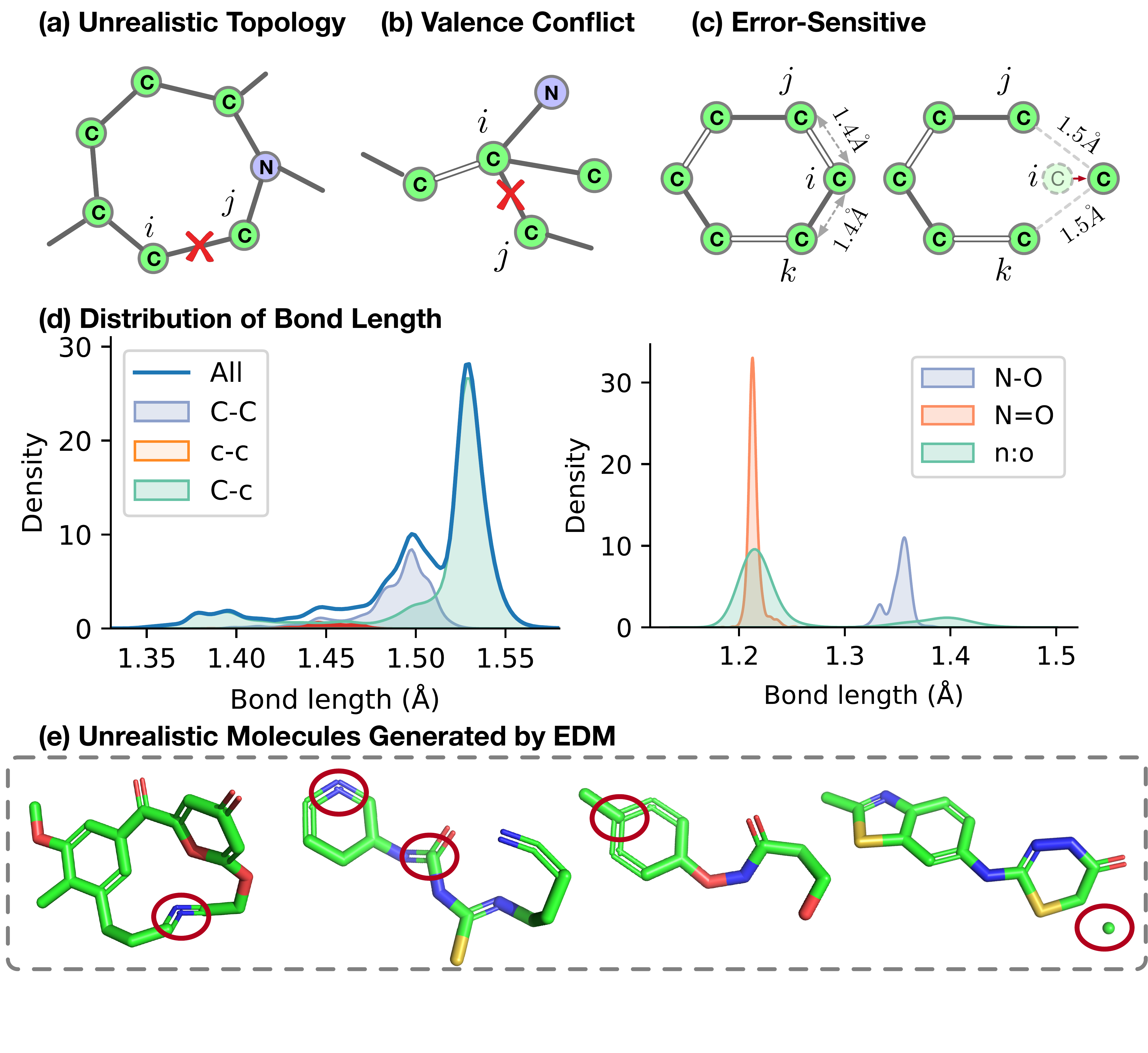}
    \vspace{-1.cm}
    \caption{The atom-bond inconsistency problem. (\textbf{a})-(\textbf{c}) Typical biases introduced by adding bonds in the post-processing manner. (\textbf{d}) Distributions of bond lengths in the GEOM-Drug subset. Left panel: lengths of single bonds connecting two carbon atoms. The upper and lower letters \textit{C} and \textit{c} represent non-aromatic and aromatic carbons, respectively. Right panel: lengths of single, double, and aromatic (symbols ``-'', ``='', and ``:'', respectively) bonds connecting nitrogen and carbon atoms. (\textbf{e}) Unrealistic 3D molecules generated by EDM.}
    \label{fig: intuition}
    \vspace{-0.5cm}
\end{figure}

The current 3D drug generation models have limited success rates in producing realistic molecules. This is mainly due to the way atoms and bonds are generated, where most models first predict atom types and positions and then add chemical bonds based on a lookup table of canonical bond lengths. However, this two-step process can introduce biases and lead to unrealistic topologies, such as extra-large ring structures that are uncommon in real molecules or violate atom valency constraints (\ref{fig: intuition}(a, b)). Additionally, ignoring bonds during training can make the model less robust to noise. For example, small errors may cause the lookup table to reject the correct aromatic bonds, as shown in Figure \ref{fig: intuition}(c), where atom $i$ should be connected to $j$ and $k$ to form a benzene ring.


Additionally, the chemical bonds are not determined solely by the distances and types of atoms. For instance, as shown in the left pane of Fig. \ref{fig: intuition}(d), even though both bonds are carbon-carbon single bonds, their lengths could be different depending on whether they connect aromatic atoms or not. The right pane of Fig. \ref{fig: intuition}(d) shows that both the nitrogen-oxygen single bond and the aromatic bond can have the same length. Another evidence is that \citet{equivariant_diffusion} found that their 3D molecule diffusion model, named EDM, made more errors for bond predictions on the drug-like dataset GEOM-Drug than on the QM9 dataset which only contains molecules with up to nine heavy atoms, indicating that post-processing bond types are difficult for large drug-like molecules. To demonstrate the influence of this sub-optimal procedure, we list multiple typically unrealistic 3D molecules generated by EDM \citep{equivariant_diffusion} in Fig. \ref{fig: intuition}(e).

To address this atom-bond inconsistency problem, we propose a new 3D molecule design model, named \model{}, which can jointly sample atoms and bonds based on a probabilistic diffusion model. The difficulty of applying the diffusion model is that even adding small noise to atoms makes the atom distances deviate from the real bond lengths so the training algorithm does not evolve a process of gradually recovering the chemical bonds from a random distribution. There might exist plenty of bonds whose lengths are far from the true lengths in the intermediate molecules during diffusion, which makes it less meaningful to learn such distributions of bond types. To solve the problem, we add different levels of noise to perturb bonds and atoms. In addition, we design a new diffusion model in which the atom generation process receives guidance from the gradient of a bond predictor such that the generated atoms are more suitable to place chemical bonds. We also propose a new E(3)-equivariant neural network in which both atom and bond representations are updated in the message passing. Finally, we find most of the previous popular metrics are inherited from 2D molecule generation tasks, which could not fully measure the qualities of generated molecules. Therefore, we also propose more metrics focusing on the drug-likeness properties, geometric properties, and rationality of generated molecules. 

\section{Related work} \label{sec:related}

\paragraph{Molecule Generation}
The most common molecular representation is 1D SMILES \citep{weininger1988smiles}, based on which many sequence-based generative models can be applied \citep{gomez2018automatic, kusner2017grammar, segler2018generating}. A more widely used representation is the 2D molecular graph thanks to the great progress in graph neural networks \citep{liu2018constrained, shi2020graphaf, jin2018junction, jin2020composing, you2018graph, zhou2019optimization}. However, the lack of spatial information makes them difficult to be applied in real scenarios where biologists want to fit the 3D structures of molecules into protein pockets. Recently, researchers start to focus on designing 3D molecules, which can be divided into two categories: \textit{autoregressive models} and \textit{non-autoregressive models}. In autoregressive models, atoms are sequentially generated based on previously generated atoms. The bonds are added using separate algorithms after all atoms are generated \cite{gebauer2019symmetry, luo2022autoregressive} or directly predicted after each atom is generated \cite{peng2022pocket2mol, roney2022generating}.
However, the intrinsic weakness of the autoregressive approach such as the unawareness of global information has hindered their generation abilities. In non-autoregressive models, molecules are represented as either atomic density grids or 3D point clouds, and VAE \citep{ ragoza2020learning}, flow \citep{satorras2021enf} or diffusion \cite{equivariant_diffusion} models are applied to generate 3D molecules. They only add bonds after all atoms have been determined and cannot include the bond information during the generation process.





\textbf{Diffusion Models for Small Molecule-Related Task}\,\, The diffusion and score-based models \citep{ho2020denoising, song2019generative, song2020score} have achieved remarkable success in image generation \citep{nichol2021improved, ramesh2022hierarchical}. They have also been applied to molecule-related tasks, such as molecular conformation generation \citep{xu2022geodiff}, \textit{de novo} molecule design \citep{equivariant_diffusion, guan2023d, schneuing2022structure, lin2022diffbp, moldiffbridge} and linker design \cite{igashov2022equivariant}. However, 3D molecules were over-simplified as point clouds in these models, that is, only atoms are considered in the diffusion model. The chemical bond information is either given as contexts or added in a post-processing manner.

\textbf{Diffusion Models for Graph Generation}\,\,
With the success of diffusion models in continuous space, a growing number of work has begun to focus on diffusion models in discrete space and more challenging data structures such as graphs. \citet{niu2020permutation} proposed the first score-based model for graph generation by thresholding a continuous value to indicate edges. Other similar works \cite{jo2022score, song2020score, vignac2022digress, haefeli2022diffusion} leverage different diffusion technologies to generate graphs. However, all of these models operate on \textbf{2D} graphs and only consider permutation invariance. Our study presents the first E(3)-equivariant diffusion model capable of producing high-quality \textbf{3D} molecular graphs and effectively tackling the atom-bond inconsistency problem. While undergoing review, we noticed another study that also aimed to generate molecular graphs with 3D conformations\cite{vignac2023midi}. However, their research focused on distinct aspects compared to ours.

\section{Method} \label{sec:method}

We first introduce the forward and reverse processes of our diffusion framework in Sec. \ref{sec: diff} and the backbone architecture of the E(3)-equivariant model in Sec. \ref{sec: network}. In Sec. \ref{sec: schedule}, we describe a new noise schedule for atom and bond diffusion during training. Finally in Sec. \ref{sec: gui}, we introduce our new bond-guided 3D molecule generation process. 

\subsection{3D Molecule Diffusion Framework} \label{sec: diff}

\textbf{Notation}\,\, A 3D small molecule is described by the atom (element) types, atom positions (coordinates in the space), and covalent chemical bonds. A molecule with $N$ atoms can be represented as $M = \{ A, R, B\}$, where $A=\{a_i\}_N \in \mathbb{A}^N$ is the atom types, $R=\{\mathbf{r}_i\}_N\in \mathbb{R}^{N\times3}$ is the atom positions, and $B=\{b_{ij}\}_{N\times N} \in \mathbb{B}^{N\times N}$ is the chemical bonds. Here $\mathbb{A}$ and $\mathbb{B}$ represent the spaces of atom types and chemical bonds, respectively. In this work, we selected seven popular element types (C, N, O, F, P, S, and Cl) and five chemical bond types including four real bond types (single, double, triple, and aromatic bonds) and one dummy type \textit{none-type}. Note that the hydrogen atoms are not considered because they can be easily inferred based on other atoms.  

\subsubsection{Forward Process}

A diffusion model defines two Markov random processes. The forward process gradually adds noise to the data based on a pre-defined noise schedule and the reverse process utilizes neural networks to remove noise and finally recover the real data from the noise. Let superscript $t$ denote the variables at time step $t$ ($t=0,1,\ldots T$), and $M^{0}$ represents the 3D molecules in the real distribution. $M^{t}$ is sampled from the distribution $q(M^{t}|M^{t-1})$ only conditioned on $M^{t-1}$. 
For atom types and bond types, since both of them are discrete, we represent them using categorical distributions. 
The forward process is formulated as follows,
\begin{equation} \label{eq: diff_one_step}
\begin{aligned}
    q(\mathbf{r}_i^{t}|\mathbf{r}_i^{t-1}) :=& \mathcal{N}(\mathbf{r}_i^{t}|\sqrt{1-\beta^{t}}\mathbf{r}_i^{t-1}, \beta^{t}\mathbf{I})  \\
    q(a_i^{t} | a_i^{t-1}) :=& \mathcal{C}( a_i^{t}|(1-\beta^{t})a_i^{t-1} + \beta^{t} \mathbbm{1}_{k}) \\
    q(b_{ij}^{t} | b_{ij}^{t-1}) :=& \mathcal{C}( b_{ij}^{t}|(1-\beta^{t})b_{ij}^{t-1} + \beta^{t} \mathbbm{1}_{k'}),
\end{aligned}
\end{equation} 
where $\beta^{t}\in [0, 1]$ is the pre-defined noise scaling schedule, $\mathbf{I}\in \mathbb{R}^{3\times3}$ is the identity matrix and $\mathbbm{1}_k$ represents a one-hot vector with one at the $k$th position and all the others zeros. 
For the atom positions we gradually add scaled standard Gaussian noise, and for atom types or bond types, we add more probability mass to the $k$th or $k'$th type so that in the forward process the atom types or bond types will gradually be perturbed to become those types. We call it \textit{absorbing} type because it works like all atom (or bond) types are gradually absorbed to this specific type \cite{hoogeboom2021argmax, austin2021structured}. For the atom type, we add one more type to the original element space as the absorbing type. For the bond type, we directly choose the none-type as the absorbing type. Note that the previous 3D molecule diffusion model EDM \cite{equivariant_diffusion} regarded the atom types as continuous vectors and perturbed them with Gaussian noise. However, using discrete values to model atom and bond types is a more natural way and also showed better performance (see results in Sec. \ref{sec: gen_abilit}). Note in Eq. \ref{eq: diff_one_step}, the pre-defined noise scaling $\beta^{t}$ for the atom types, atom positions, and bond types can be different but we do not differentiate them in the equation for clarity.

By leveraging the Markov property, $M^{t}$ could be directly derived from the original sample $M^{0}$ (i.e., $q(M^{t}|M^{0})$) based on Eq. \ref{eq: diff_one_step}. If we define $\alpha^t := 1 - \beta^t$ and $\bar{\alpha}^t := \prod_{s=1}^t \alpha^s$, the sample $M^{t}$ can be derived as:
\begin{equation} \label{eq: diff_t}
\begin{aligned}
    q(\mathbf{r}_i^{t} | M^{0}) &= \mathcal{N}( \mathbf{r}_i^{t} | \sqrt{\bar{\alpha}^t}\mathbf{r}_i^{0}, (1 - \bar{\alpha}^t) \mathbf{I})  \\
    q({a}_i^{t} | M^{0}) &= \mathcal{C}( {a}_i^{t} | \bar{\alpha}^t {a}_i^{0} + (1 - \bar{\alpha}^t) \mathbbm{1}_{k} ) \\
    q({b}_{ij}^{t} | M^{0}) &= \mathcal{C}( {b}_{ij}^{t} | \bar{\alpha}^t {b}_{ij}^{0} + (1 - \bar{\alpha}^t) \mathbbm{1}_{k'} ).
\end{aligned}
\end{equation}
From this equation, it is obvious that the values of $\bar{\alpha}^t$ can be interpreted as how much information from the real data is still preserved at step $t$. We call $\bar{\alpha}^t$ \textit{Information level} and it can be determined by the noise level $\beta^t$ (we discuss how to choose them in Sec. \ref{sec: schedule}).

As $t\rightarrow \infty$, we get $q({a}_i^{t} | M^{0})\rightarrow \mathbbm{1}_{k}$,\; $q(\mathbf{r}_i^{t} | M^{0}) \rightarrow \mathcal{N}(\mathbf{0}, \mathbf{I})$, and $q({b}_{ij}^{t} | M^{0})\rightarrow \mathbbm{1}_{k'}$ according to Eq. \ref{eq: diff_t}, indicating that for large $T$, the atom positions approximately follow the standard Gaussian distribution, and the atom and bond types place all the probability mass on the absorbing types for $t=T$. These distributions, called prior distributions, will serve as the starting distributions of the reverse process.



\subsubsection{Reverse Process}
In the reverse process, we reverse the Markov chain to reconstruct the true sample from prior distributions and use E(3)-equivariant neural networks to parameterize the transition $p_{\theta}(M^{t-1}|M^{t})$.
Specifically, we model the predicted atom positions as Gaussian distribution $\mathcal{N}(\mathbf{r}_i^{t-1} | \boldsymbol{\mu}_\theta(M^{t}, t), \Sigma^{t})$, the atom and bond type as Categorical distributions $\mathcal{C}(a_i^{t-1}|H_\theta(M^{t}, t))$ and $\mathcal{C}(b_{ij}^{t-1}|H'_\theta(M^{t}, t))$, respectively. Here $\boldsymbol{\mu}_\theta, {H}_\theta$ and ${H}'_\theta$ are all neural networks and $\Sigma^{t}$ is set as $\beta^{t}$. During training, we add noise to data and train the neural network to recover $M^{t-1}$ from $M^{t}$ by optimizing the predicted distributions $p_{\theta}(M^{t-1}|M^{t})$ to approximate the true posterior $q(M^{t-1}|M^{t}, M^{0})$ (the true posterior can be derived from Eq. \ref{eq: diff_one_step} and \ref{eq: diff_t}, see Appendix \ref{sec: app_diff}). Here both $M^{t-1}$ and $M^{t}$ are collected in the forward process. The loss function is defined as follows,
\begin{equation}
\begin{aligned}
\nonumber
    L^{t-1} &= L_{\text{pos}}^{t-1} + \lambda_1 L_{\text{atom}}^{t-1} + \lambda_2 L_{\text{bond}}^{t-1} \\
\end{aligned}   
\end{equation}

\begin{equation}
\begin{aligned}
\nonumber
    L_{\text{pos}}^{t-1} &= \frac{1}{N}\sum_i \| \mathbf{r}_i^{t-1} - \boldsymbol{\mu}_\theta(M^{t},t)_i \|_2^2 \\
    L_{\text{atom}}^{t-1} &= \frac{1}{N} \sum_i  D_{\text{KL}}[q(a_i^{t-1} | M^{t}, M^{0}) \| p_\theta(a_i^{t-1}|M^{t})]  \\
    L_{\text{bond}}^{t-1} &= \frac{1}{N^2} \sum_{ij}  D_{\text{KL}}[q(b_{ij}^{t-1} | M^{t}, M^{0}) \| p_\theta(b_{ij}^{t-1}|M^{t})],
\end{aligned}   
\end{equation}
where $\lambda_1$ and $\lambda_2$ are pre-defined constants. During training, we randomly sample a step $t$ and optimize the neural network by minimizing the loss $L^{t-1}$. After the training process, to generate new molecules, we first sample $M^{T}$ from the prior distributions $p(M^{T})$ and then repeatedly sample from $p_{\theta}(M^{t-1}|M^{t})$ for $t=T,T-1,\ldots,1$ to gradually remove noise. The prior distribution $p(M^{T})$ is the standard Gaussian distribution $\mathcal{N}(\mathbf{0}, \mathbf{I})$ for atom positions and the Categorical distributions with all probability mass on the absorbing types for atom types and bond types. 

\subsection{Equivariant Graph Neural Networks}  \label{sec: network}
An important property the neural network should possess for modeling 3D molecules is E(3)-equivariance, i.e., the outputs of the network should be equivariant to any 3D rotation, translation, and reflection. The previous model EDM utilized the E(3)-equivariant network EGNN \cite{satorras2021egnn} to update atom representations by passing messages between atoms. However, as we focus on the atom-bond inconsistency problem, we need to design a new equivariant network architecture and message passing algorithms involving both atoms and bonds. Formally, for an input molecule $M=\{ \{a_i\}_N, \{\mathbf{r}_i\}_N, \{b_{ij}\}_{N\times N} \}$, we construct a complete graph with coordinates in which vertices are the atoms and all vertices are connected. We use notations 
$\mathbf{v}_i\in\mathbb{R}^{d}$ and $\mathbf{e}_{ij}\in\mathbb{R}^{d'}$ to denote the hidden representations for vertex $i$ and edge $\langle i,j\rangle$, respectively. The input vertex features are the one-hot encoding of atom types and the input edge features are the one-hot encoding of bond types. Then the updating of vectors $\mathbf{v}_i,\mathbf{e}_{ij}$ and coordinates $\mathbf{r}_i$ is defined as follows,
\begin{equation} \label{eq: model}
\begin{aligned}
    \tilde{\mathbf{e}}_{ij} &\leftarrow \phi_d(\mathbf{e}_{ij}, \|\mathbf{r}_i - \mathbf{r}_j\|_2) \\
    \mathbf{v}_i &\leftarrow \text{Linear}(\mathbf{v}_i) + \sum_j \phi_v(\mathbf{v}_j, \tilde{\mathbf{e}}_{ij}, t)  \\
    \mathbf{e}_{ij} &\leftarrow \sum_k \phi_e (\mathbf{v}_k, \tilde{\mathbf{e}}_{ki}, t) + \sum_k \phi_e (\mathbf{v}_k, \tilde{\mathbf{e}}_{jk}, t) \\
     &\quad +\text{Linear}(\mathbf{v}_i) + \text{Linear}(\mathbf{v}_j) + \text{Linear}(\tilde{\mathbf{e}}_{ij}) \\
    \mathbf{r}_i &\leftarrow \mathbf{r}_i + \sum_{j} \phi_r(\mathbf{v}_i, \mathbf{v}_j, \tilde{\mathbf{e}}_{ij}, t) \frac{\mathbf{r}_i - \mathbf{r}_j}{\|\mathbf{r}_i - \mathbf{r}_j\|_2^2},
\end{aligned}
\end{equation}
where $\text{Linear}(\cdot)$ represents linear transformations of the inputs, and $\phi_d, \phi_v, \phi_e, \phi_r$ are neural networks made up of different multilayer perceptrons (MLPs) (the detailed architectures can be found in Appendix \ref{sec: app_model}). The main difference between the proposed model and the previous models (e.g., EGNN) is that we address the importance of edges. The edges also propagate messages with other edges and nodes to update their representations. This model is specifically designed for this diffusion framework where the features of both nodes and edges are used to predict the atom and bond types. After multiple rounds of updates, we take the coordinates $\mathbf{r}_i$ as the predicted mean of the atom positions $\boldsymbol{\mu}_\theta$ and apply another two MLPs followed by softmax activations to translate the feature vectors $\mathbf{v}_i$ and $\mathbf{e}_{ij}$ to probabilities of atom and bond types as:
\begin{equation}
\begin{aligned}
p(a_i^{t-1}|M^{t}) &= \mathcal{C}(a_i |\text{softmax}(\text{MLP}(\mathbf{v}_i))) \\
p(b_{ij}^{t-1}|M^{t}) &= \mathcal{C}(b_{ij} |\text{softmax}(\text{MLP}(\mathbf{e}_{ij} + \mathbf{e}_{ji}))). \\
\end{aligned}
\end{equation}
Note that the bond type $b_{ij}$ is determined by both the edge features $\mathbf{e}_{ij}$ and $\mathbf{e}_{ji}$ for symmetry. 

\subsection{The Atom and Bond Noise Schedule}  \label{sec: schedule}

There is a unique factor in the diffusion generation of 3D molecules. Since the bond types of the molecules have a strong relationship with the atom distances and atom types, if the bond types use the same noise schedules as the atom types and the atom positions, the noised data distribution will suffer from the inconsistency of atoms and bonds. More specifically, as more noise is added to the molecule, the distances of atom pairs that originally form chemical bonds will easily deviate from the bond lengths and there is no need to further perturb the bonds. For instance, if two carbon atoms form a single bond in a molecule, during the diffusion process it is quite possible that their distance becomes larger than $3\rm\AA$ which cannot form any chemical bond. In this case, it is meaningless to ask the neural network to learn that two atoms that are so far away from each other can form a single bond. More formally, the distribution of real molecules $p(M)=p(A,R,B)$ can be decomposed as $p(M)=p(A,R)p(B|A,R)$. For any diffusion step $t$, the conditional distribution $p(B^{t}|A^{t}, R^{t})$ could greatly shift from the true distribution $p(B^{0}|A^{0}, R^{0})$. Therefore, the intermediate molecules $M^{t}=(A^{t}, R^{t}, B^{t})$ contains biased relationship between atoms and bonds.

To address this problem, we propose a new diffusion strategy by separating the diffusion processes of bonds and atoms and dividing the forward diffusion process into two stages. In the first stage, we make the bond types diffuse to the prior distribution and all bonds gradually become the absorbing none-bond type. In this stage, the atom types and positions are only slightly perturbed. Note that it is important to inject little noise to atom types and positions instead of fixing all of them and only learning to recover the bonds. In this way, the model becomes more robust and can acquire the knowledge that a chemical bond needs to be added when the distance of two atoms is within a certain range instead of a fixed value. At the next timestamp, the model can also learn to adjust the coordinates of atoms to satisfy the chemical constraints based on the bond just recovered. At the end of the first stage, almost all bonds are labeled as none-type and reach the prior distribution. 
In the second stage, we keep perturbing the atom types and positions to arrive at their corresponding prior distributions. Since the atom positions have greatly changed in this stage, it is reasonable that no real chemical bonds still exist in the molecules and thus the model focuses on the learning of atom types and positions. Overall, this strategy keeps the relationship between the bonds and atoms more similar to the true one during diffusion. 

\begin{figure}
    \centering
    \includegraphics[width=\linewidth]{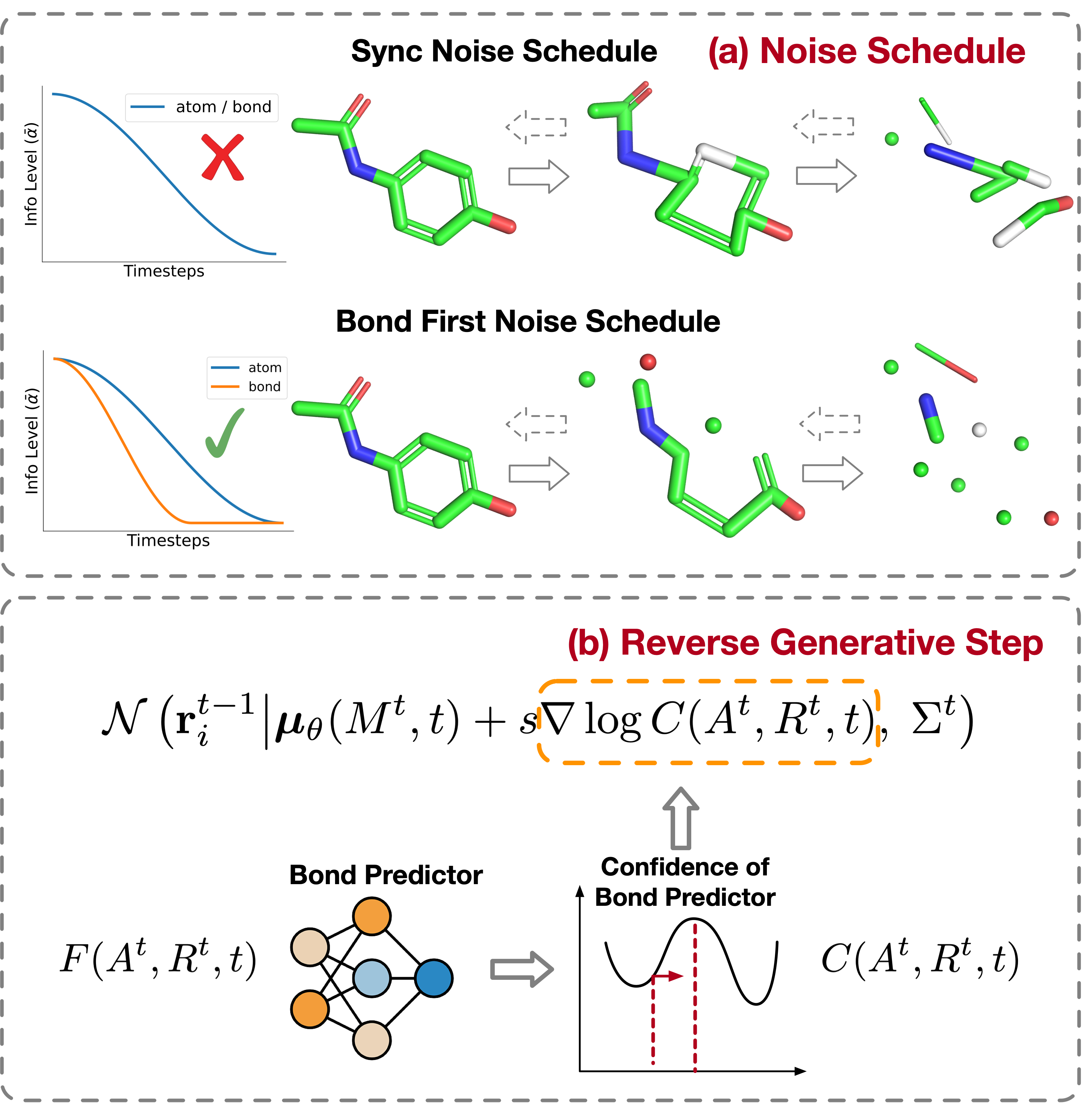}
    \caption{Model Illustration. (\textbf{a}) The curves of the information level $\bar{\alpha}_t$ w.r.t $t$ and an illustration of the diffusion process for the original simple (synchronized) noise schedule and our proposed (bond first) noise schedule. (\textbf{b}) The guidance of the bond predictor for the generation step.}
    \label{fig: model}
    \vspace{-0.5cm}
\end{figure}

We implement this by designing a new noise schedule for the bonds and atoms. As shown in Fig. \ref{fig: model}(a), we set different $\beta^{t}$ for the atoms (types and positions) and bond types so that the information level $\bar{\alpha}^{t}$ of bond types decay to zeros much faster than the atoms during the diffusion process. In the first stage, the atoms are only marginally perturbed and the model pays more attention to denoise the bond types. In the second stage, almost all real bonds have been removed and the model only focuses on the prediction of atoms. In this way, the model does not need to learn the bond types when the atom distances have obviously deviated from the canonical bond lengths. More details can be found in Appendix \ref{sec: app_noise}.

\subsection{Guidance of Bond Predictor}  \label{sec: gui}

As shown in Fig. \ref{fig: intuition}(d), the lengths of real chemical bonds are within a very small range. For instance, the nitrogen-oxygen single bonds (N-O) have a canonical length around $1.35\rm\AA$ and seldom lower than $1.31\rm\AA$ or greater than $1.39\rm\AA$. If the generative model happens to generate an N-O atom pair with distance $1.25\rm\AA$, the model should fix the atom positions so that the bond length falls into the correct distance range of N-O bonds. 
Therefore, we can also take advantage of the strong relationship between the bond types and bond lengths to guide the generation of atom positions so that our model could place the atoms in the correct positions and thus generate accurate 3D structures of the molecules. 

To achieve this intuition, we train another equivariant graph neural network $F({A}^t,{R}^t, t)$ which takes the atom types and positions as input and predicts all the bonds for the molecule, i.e., a neural-network-based bond predictor. Assume the predicted logits of the bond types for bond $\langle i, j\rangle$ are $F({A}^t,{R}^t, t)_{ij}$, and we define a function $c[F({A}^t,{R}^t, t)_{ij}]$ to quantify the confidence of the predictions for bond $\langle i, j\rangle$. The confidence for all bonds based on atom sets $\{A, R\}$ is
\begin{equation}
    C({A}^t,{R}^t, t) = \prod_{ij} c[F({A}^t,{R}^t,t)_{ij}].
\end{equation}
If the sampled atom positions are not accurate, the confidence of the bond predictor $C({A}^t,{R}^t,t)$ will be low, and vice versa. 
Then the gradient of the $\log C({A}^t,{R}^t,t)$ with respect to $R^t$ provides us the direction to improve the confidence and thus guide the determination of atom positions. Formally, with the guidance, the reverse generation of the atom positions $p_\theta(\mathbf{r}_i^{t-1}|M^t)$ can be revised as:
\begin{equation}  \label{eq: gui}
    \mathcal{N}\left(\mathbf{r}_i^{t-1} \big| \boldsymbol{\mu}_\theta(M^t, t) + s\nabla \log C(A^t,R^t, t), \; \Sigma^t\right),
\end{equation}
where $s$ is a hyper-parameter and set as $1\times 10^{-4}$. The intuition comes from the classifier guidance in diffusion generative models \citep{dhariwal2021diffusion}, where a classifier is utilized to guide the generated samples to optimize the specific property. But here we are not using the bond predictor to improve some specific property but we want to improve the confidence of the bond predictor which is equivalent to rendering more accurate bond lengths and thus more accurate atom positions. The bond predictor $F(A^t, R^t, t)$ needs to be separately trained. To adapt it for the noised molecules, we train the bond predictor which takes the noised atom types $A^t$, noised atom positions $R^t$, and step $t$ as input and then predicts the ground truth bond types $B^0$.  More details about the bond predictor are shown in Appendix \ref{sec: app_bondpred}.

Now we discuss the choice of confidence function $c[F({A}^t,{R}^t,t)_{ij}]$. If the outputs of a classifier are logits, the exponential logits of the Categorical distribution can be regarded as the parameters of its conjugate prior distribution Dirichlet distribution \citep{uncertainty}. Therefore, the uncertainty of the prediction can be derived from logits and the confidence function is defined as the reciprocal of the uncertainty. If we represent $F({A}^t,{R}^t,t)_{ij}$ as $\mathbf{y}$, the confidence function is:
\begin{equation}
    c(\mathbf{y}) = \sum_k\exp(y_k) +1.
\end{equation}
where $y_k$ is the logit for the $k$th type. 

\section{Results} \label{sec: results}

\begin{table*}[t]
    \vspace{-0.2cm}
    \centering
    \caption{The generation abilities of EDM, \model{}, and variants of \model{}}
    \begin{tabular}{l|ccc|cccc} 
    \toprule
        ~ & Validity & Connectivity & Succ. rate & Novelty & Diversity & Uniqueness & Sim. Val.  \\ \hline
        EDM & 0.447  & 0.830  & 0.371  & \textbf{1.000}  & 0.729  & \textbf{1.000}  & 0.441   \\ 
        MolDiff & \textbf{0.997}  & \textbf{0.996}  & \textbf{0.993}  & 0.972  & \textbf{0.769}  & 0.986  & \textbf{0.634}   \\ \hline
        Predict bond (Lookup table) & 0.297  & 0.937  & 0.278  & \textbf{1.000}  & 0.751  & 0.999  & 0.434   \\ 
        Predict bond (NN Predictor) & 0.956  & 0.978  & 0.935  & 0.984  & 0.768  & 0.999  & 0.592   \\ 
        Predict bond (Openbabel) & 0.992  & 0.969  & 0.961  & 0.986  & 0.762  & 0.999  & 0.576   \\ 
        Conti. diffusion  & 0.914  & 0.885  & 0.808  & 0.999  & 0.752  & \textbf{1.000}  & 0.513   \\ 
        Conti. diffusion (scaling) & 0.972  & 0.934  & 0.907  & 0.997  & 0.759  & \textbf{1.000}  & 0.511   \\ 
        Add bond length loss & 0.998  & 0.959  & 0.958  & 0.986  & 0.752  & 0.998  & 0.590  \\
    \bottomrule
    \end{tabular}
    \label{tab: gen_ability}
    \vspace{-0.5cm}
\end{table*}

We first compare the generation abilities of \model{} with multiple baseline models in Sec. \ref{sec: gen_abilit}. Then in Sec. \ref{sec: drug_prop}, we expand the comparison to multiple new metrics which are usually neglected in previous 3D molecule generation tasks. Finally, we analyze the effectiveness of our design through an ablation study (Sec. \ref{sec: ablation}) and model analysis (Sec. \ref{sec: model_ana}) to explore the contribution of each component.

\subsection{Generation Ability} \label{sec: gen_abilit}

\textbf{Dataset}\,\,  We utilized the GEOM-Drug dataset to train and assess our models, and included details about the data pre-processing in Appendix \ref{sec: app_data}. To better reflect the real drug design scenario, we only considered molecules with major elements (C, N, O, F, P, S, and Cl) and excluded hydrogen atoms. However, we also conducted additional evaluations that accounted for minor element types (B, Br, I, Si, and Bi) or still kept hydrogen atoms, and additionally, we made a comparison with more baselines in another widely used dataset QM9, which are further elaborated in Appendix \ref{sec: more_eval}.


\textbf{Baselines and Setup}\,\, We compared \model{} with EDM and multiple variants of our \model{} model. EDM is the first diffusion model for 3D molecule generation, which applied continuous diffusion for atom types and positions and added bond types using a lookup table in a post-processing manner. To highlight the importance of bonds in the diffusion model, we provided three variants of \model{} that did not involve any bond information during diffusion and predicted bonds using three different strategies after generating the atoms, including 1) using the lookup table (same as EDM), 2) training another neural network as the bond predictor, and 3) adopting the chemical toolbox Open Babel \citep{o2011obabel}. 
We provided another two variants of \model{} that used continuous embedding to encode atom and bond types to demonstrate the performance difference against using discrete embedding. 
One of the variants implemented the strategy of EDM which added additional scaling to the continuous features \cite{equivariant_diffusion}. Moreover, we provided another baseline model which added an additional loss to minimize the real bond lengths of the predicted one and the ground truth, a seemingly natural and simple way to generate better bond lengths. For each model, we sampled 1000 valid and complete 3D molecules and repeated this procedure for three times. We reported the average of the metrics in the main text and postponed the standard deviations in Appendix \ref{sec: app_std}. More details about the settings of these baselines can be found in Appendix \ref{sec: app_baseline}. 

\textbf{Metrics}\,\, We defined three metrics to evaluate the basic learning abilities of the models: (1) \textbf{Validity} is the ratio of valid molecules (can be parsed by RDKit) among all generated molecules; (2) \textbf{Connectivity} is the ratio of complete molecules (in which all atoms should be connected) among all valid molecules; (3) \textbf{Success rate} is the ratio of valid and complete molecules among all generated molecules, which is equivalent to validity multiplied by connectivity. These three metrics are the most important ones because they measure whether the models learn basic chemical rules such as valence constraint and completeness of molecules. Next, among the valid and complete molecules, we measure the similarities from four different aspects: (1) \textbf{Novelty} is the ratio of generated molecules that do not exist in the training set; (2) \textbf{Diversity} is the average molecular fingerprint similarities of all pairs of generated molecules; (3) \textbf{Uniqueness} is the ratio of unique molecules among all generated ones; (4) \textbf{Similarity with validation set} is the average similarities between the generated molecules and the validation set.

The comparison results are shown in Tab. \ref{tab: gen_ability}. For the validity, connectivity, and success rate, \model{} outperformed EDM by a large margin, indicating that \model{} better captured the chemical constraints of the 3D molecules. Furthermore, the success rate of \model{} approximated to one, suggesting that almost all generated molecules were complete and thus \model{} are ready to be applied to real downstream tasks. In comparison with other variants of \model{}, we found the exclusion of bonds during diffusion harmed the learning and generation abilities of the models. Tab. \ref{tab: gen_ability} also suggested that the diffusion of atom and bond types in discrete space achieved better performance than in continuous space. Adding bond lengths as an additional loss did not improve the performance either. One possible reason is that it might harm the learning of the noised positions during diffusion.

Among the valid molecules, \model{} exceeded EDM and all variants for the diversity and similarity with the validation set, indicating that \model{} generated not only more diverse molecules but also more similar ones to the real molecules. As for novelty and uniqueness, \model{} did not achieve the best values but the values were good enough since both of them approximated to one. Overall, we concluded that \model{} had much better learning and generation ability than EDM, and the inclusion of bonds during diffusion and treating atom/bond types as discrete variables were both necessary components for the generation of 3D molecules.

\subsection{Quality of 3D Molecules} \label{sec: drug_prop}

The above metrics only focused on whether the generated molecules are reasonable or diverse. Here we dived into the quality related to 3D geometries and chemical properties of the molecules. These metrics are significant for a sampled molecule to be further selected as a drug candidate and used for real-world applications. We analyzed the valid and complete molecules of EDM and \model{} from four aspects: drug-likeness, 3D structures, bonds, and rings. 

\begin{table}[ht]
    \vspace{-0.5cm}
    \centering
    \caption{The qualities of the generated 3D molecules}
    \begin{tabular}{l|l|p{0.5cm}c}
    \toprule
        Group & Metrics  & EDM & MolDiff  \\ \hline
        \multirow{3}*{\makecell[l]{Drug-\\likeness}} & QED  (↑) & 0.558  & \textbf{0.668}   \\ 
        ~ & SA  (↑) & 0.568  & \textbf{0.874}   \\
        ~ & Lipinski  (↑) & 4.923  & \textbf{4.986}   \\ \hline
        \multirow{4}*{\makecell[l]{3D\\structures}} & RMSD  (↓) & 1.321  & \textbf{0.939}   \\ 
        ~ & JS. bond lengths  (↓) & \textbf{0.246}  & 0.365   \\ 
        ~ & JS. bond angles  (↓) & 0.282  & \textbf{0.155}   \\ 
        ~ & JS. dihedral angles  (↓) & 0.328  & \textbf{0.162}   \\ \hline
        \multirow{5}*{Bonds} & JS. \#bonds per atoms  (↓) & 0.139  & \textbf{0.115}   \\ 
        ~ & JS. basic bond types  (↓) & 0.306  & \textbf{0.093}   \\ 
        ~ & JS. freq. bond types  (↓) & 0.378  & \textbf{0.163}   \\ 
        ~ & JS. freq. bond pairs  (↓) & 0.396  & \textbf{0.136}   \\ 
        ~ & JS. freq. bond triplets  (↓) & 0.449  & \textbf{0.125}   \\ \hline
        \multirow{3}*{Rings} & JS. \#rings  (↓) & 0.106  & \textbf{0.062}   \\ 
        ~ & JS. \#n-sized rings  (↓) & 0.107  & \textbf{0.092}   \\ 
        ~ & \#Intersecting rings   (↑) & 3.667  & \textbf{8.000}  \\ 
    \bottomrule
    \end{tabular}
    \label{tab: drug_prop}
\vspace{-0.5cm}
\end{table}

\begin{figure}[htbp]
    \centering
    \includegraphics[width=\linewidth]{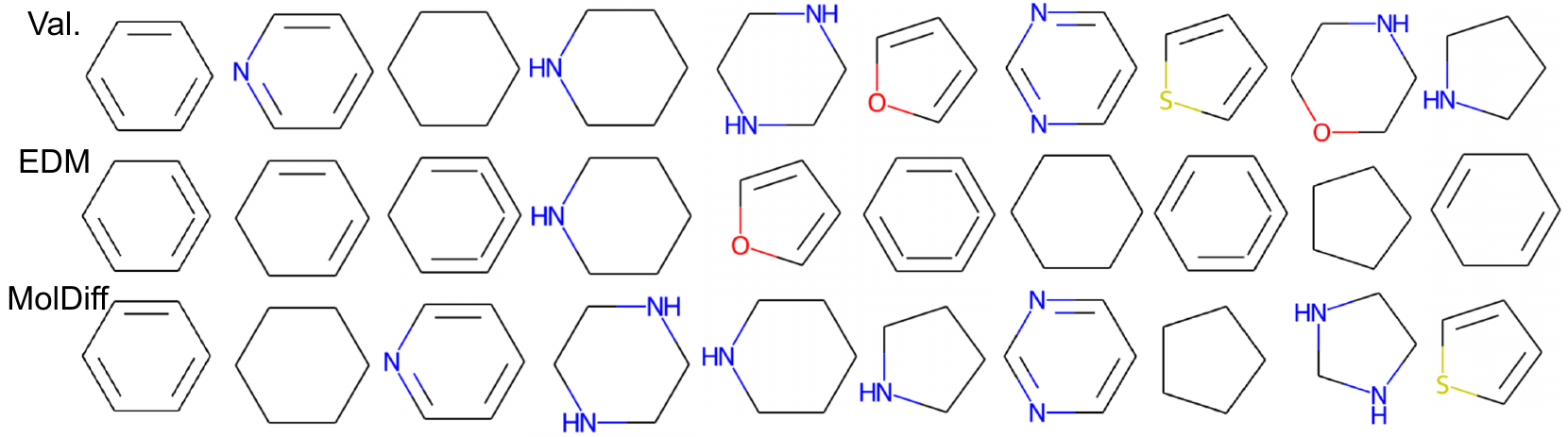}
    \caption{The most frequent ten rings among molecules in the validation set, generated by EDM, and generated by \model{}.}
    \label{fig: ring}
\end{figure}

\textbf{Drug-likeness}\,\, We evaluated drug-likeness of the generated molecule using the following metrics: (1) \textbf{QED} is the abbreviation of quantitative estimation of drug-likeness; (2) \textbf{SA} represents the synthetic accessibility score (higher values indicate easier for drug synthesis); (3) \textbf{Lipinski} measures how many rules the molecule follows the Lipinski's rule of five (details in Appendix \ref{sec: app_metric}). As shown in Tab. \ref{tab: drug_prop}, \model{} outperformed EDM by a large margin, suggesting that the molecules of \model{} were better candidates for drugs. 

\textbf{3D structures}\,\, A main difference between the 3D generation and 2D molecule graph generation is the determination of atom positions, and thus it is necessary to measure their accuracy. First, we calculated the minimal RMSDs between the generated 3D molecules and 100 possible conformations predicted by the RDKit toolkit. Next, we selected the most frequent bonds, bond pairs, and bond triplets in the validation set (the exact types are listed in Appendix \ref{sec: app_metric}). We calculated the bond lengths, bond angles, and dihedral angles for the generated molecules and the ones in the validation set. We then used the Jensen-Shannon(JS) divergence to measure the differences in the distributions between the generated molecules and the validation set. As shown in Tab. \ref{tab: drug_prop}, \model{} had much lower RMSDs, JS divergence of bond angles, and dihedral angles, indicating that the 3D structures of molecules generated by \model{} were much closer to the ground truth. We also noticed that the \model{} performed worse in terms of JS divergence of bond lengths. This was not surprising because EDM added bonds by comparing the bond lengths with a pre-defined lookup table and thus their lengths were closer to the canonical lengths. But overall, \model{} better captured the entanglement of 3D information of molecules than EDM.

\textbf{Bonds}\,\, As we particularly addressed the bond information during generation, here we analyzed the bond-related properties of the generated molecules. First, we analyzed whether models generated excessive or insufficient bonds by comparing the distributions of counts of bonds per atom between generated molecules and the validation set. Next, we analyzed the distributions of different bond types, including the basic bond types (single, double, triple, and aromatic bonds) and frequent bond types, bond pairs, and bond triplets that were used in the evaluation of 3D structure. As shown in Tab. \ref{tab: drug_prop}, \model{} showed lower JS divergence than EDM for all the distributions, which indicated that the molecules of \model{} not only had more realistic counts but also exhibited more balanced ratios among different bonds.

\begin{table}[tbp]
    \vspace{-0.5cm}
    \centering
    \caption{Ablation study (metrics of 3D structures, bonds and rings).}
    \begin{tabular}{p{{2.95cm}}|p{0.8cm}p{0.8cm}p{0.8cm}p{0.8cm}}
    \toprule
        Metrics & \makecell[l]{Mol-\\Diff} & \makecell[l]{No\\Gui.} & \makecell[l]{No\\Sche.} & \makecell[l]{Neither} \\ \hline
        RMSD & \textbf{0.939}  & 1.027  & 1.005  & 1.126   \\ 
        JS. bond lengths & \textbf{0.365}  & 0.378  & 0.461  & 0.468   \\ 
        JS. bond angles & \textbf{0.155}  & 0.162  & 0.224  & 0.209   \\ 
        JS. dihedral angles & \textbf{0.162}  & 0.202  & 0.177  & 0.204   \\ \hline
        JS. \#bonds per atoms & \textbf{0.115}  & 0.154  & 0.219  & 0.196   \\ 
        JS. basic bond types & 0.093  & 0.054  & \textbf{0.039}  & 0.087   \\ 
        JS. freq. bond types & 0.163  & \textbf{0.148}  & 0.161  & 0.184   \\ 
        JS. freq. bond pairs & 0.136  & \textbf{0.107}  & \textbf{0.107}  & 0.141   \\ 
        JS. freq. bond triplets & 0.125  & \textbf{0.102}  & 0.112  & 0.150   \\ \hline
        JS. \#rings & \textbf{0.062}  & 0.105  & 0.176  & 0.159   \\ 
        JS. \#n-sized rings & \textbf{0.092}  & 0.102  & \textbf{0.092}  & 0.119   \\ 
        \#Intersecting rings & \textbf{8.000}  & 7.667  & 6.000  & 6.000  \\ 
    \bottomrule
    \end{tabular}
    \label{tab: ablation}
    \vspace{-0.1cm}
\end{table}

\textbf{Rings}\,\, An erroneous generation of bonds will violate the graph topology of the molecules and easily result in biased ring distributions. For instance, adding unnecessary bonds will lead to redundant or unrealistic rings. We thus further extended the analysis of bonds to rings. First, we compare the distributions of counts of rings in each molecule between generated molecules and the validation set. Next, we compared the distributions of counts of $n$-sized rings between generated molecules and validation set using JS divergence and then averaged all JS divergence for $n\in\{3, 4, \ldots, 9\}$. Real molecules usually have plenty of 6-sized rings but much fewer large or small rings and thus we want to check whether the model learned such underlying design principles. Finally, we found out the most frequent ten types of rings in the validation set and generated molecules respectively, and calculated the numbers of frequent rings existing in both the validation set and the generated molecules. As shown in Tab. \ref{tab: drug_prop}, \model{} produced better distributions of counts of rings and $n$-sized rings. Regarding the ring types, \model{} generated much more realistic ones. We also displayed the most frequent ten rings among molecules in the validation set, generated by EDM and by \model{} in Fig. \ref{fig: ring}, where it is obvious that most rings generated by EDM were unrealistic as a result of erroneous bonds while those generated by \model{} were similar with the validation set. 

\subsection{Analyses of the Noise Schedule and the Guidance}  \label{sec: ablation}

To study the importance of different components, we conducted an ablation study to show the effectiveness of the guidance and the special noise schedule. We compared the complete \model{} with three variants: (1) the model without the guidance of the bond predictor; (2) the model without using the newly proposed noise schedule, and 3) the model with neither of these two techniques. Tab. \ref{tab: ablation} demonstrated that both the guidance and new noise schedule could greatly boost the determination of 3D structures. The two strategies also contributed to more realistic counts of bonds. We also found that they had no effect on balancing different bond types. Another observation was that both strategies can benefit the arrangement of the count of rings and different ring types. Therefore, we conclude that the guidance of the bond predictor and the new noise schedule are necessary components to generate molecules with better 3D structures and distributions of bonds and rings. In addition, we found these two strategies did not harm the basic metrics such as generation abilities and drug-likeness as shown in Appendix \ref{sec: app_ablation}.

\subsection{Model Analysis} \label{sec: model_ana}

\begin{figure}
    \centering
   \includegraphics[width=\linewidth]{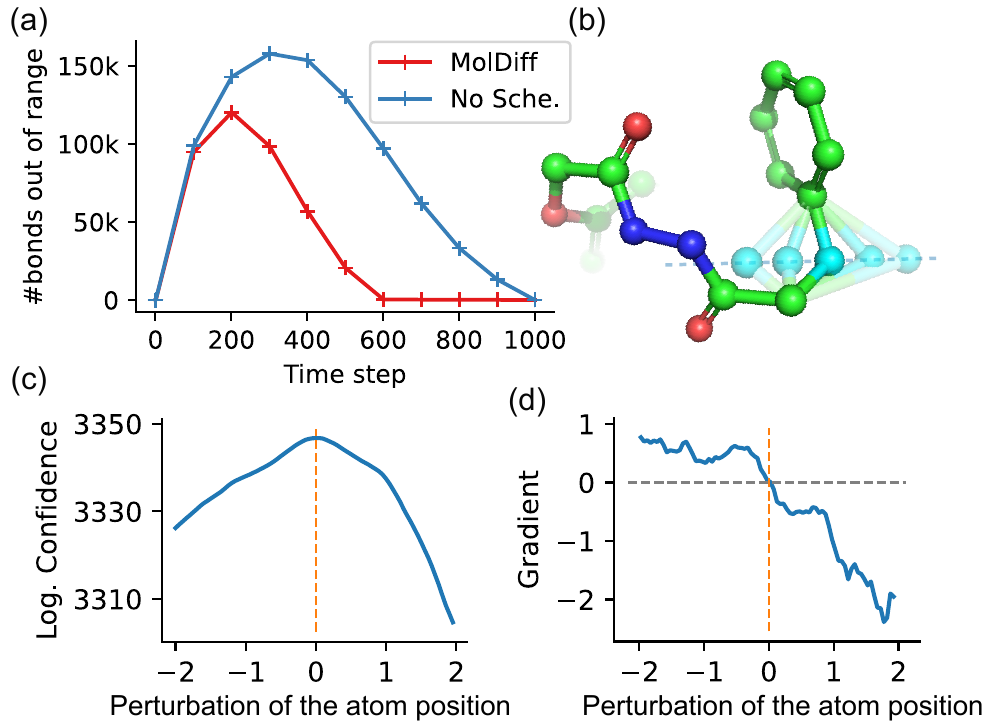}
    \caption{Model Analysis. (\textbf{a}) The numbers of bonds out of range for different noise schedules. (\textbf{b})-(\textbf{d}) An example demonstrating how the confidence function and gradient change as an atom (shown in cyan) is perturbed.}
    \label{fig: model_ana}
    \vspace{-0.5cm}
\end{figure}

Finally, we explored how the noise schedule and the guidance worked. As we mentioned above, the new noise schedule diffused the bond types earlier to avoid the irrational relationship between bonds and atoms during diffusion. We first calculated the ranges of bond lengths of real molecules in the validation set and then counted real bonds whose lengths were out of the range at different steps during the diffusion process. As shown in Fig. \ref{fig: model_ana}(a), the noise schedule of \model{} induced much fewer bonds that were out of range. Actually, these bonds were unnecessary to learn at the current moment, and thus the noise schedule of \model{} enabled the model to focus on more meaningful data distributions. As for the guidance of the bond predictor, we showed an example demonstrating how it guided the generation of positions in Fig. \ref{fig: model_ana}(b)-(d). We perturbed the position of one atom of a molecule in the dataset along the $y$-axis (Fig. \ref{fig: model_ana}(b)) and calculated the values $\log C(A, R, 0)$ and their scaled gradients w.r.t perturbations (Fig. \ref{fig: model_ana}(c,d)). We found that as the atom deviated from the ground truth position, the confidence values decreased and the gradients also deviated from zero. Therefore, the gradients of the confidence function assisted to push the atoms to the right positions. In conclusion, both the schedule and guidance facilitate \model{} to better learn and generate molecules with more accurate 3D structures and more realistic distributions of bonds and rings.

\section{Conclusion}\label{sec:conclusion}

In this work, we propose a 3D molecule generation model to simultaneously address the generation of atoms and bonds of molecules. Experimental results demonstrate that the model has greatly improved the generation ability and the quality of the generated molecules. Further work can explore more sophisticated noise schedules and more advanced guidance strategies, and analyze its potential for downstream real tasks.

\section*{Acknowledgement}

We thank Shitong Luo and Xingchao Liu for the useful discussions. We thank the anonymous reviewers for their insightful reviews. This work is supported by the National Key R\&D Program of China No. 2021YFF1201600.


\bibliography{ref}

\begin{thebibliography}{44}
\providecommand{\natexlab}[1]{#1}
\providecommand{\url}[1]{\texttt{#1}}
\expandafter\ifx\csname urlstyle\endcsname\relax
  \providecommand{\doi}[1]{doi: #1}\else
  \providecommand{\doi}{doi: \begingroup \urlstyle{rm}\Url}\fi

\bibitem[Austin et~al.(2021)Austin, Johnson, Ho, Tarlow, and van~den
  Berg]{austin2021structured}
Austin, J., Johnson, D.~D., Ho, J., Tarlow, D., and van~den Berg, R.
\newblock Structured denoising diffusion models in discrete state-spaces.
\newblock \emph{Advances in Neural Information Processing Systems},
  34:\penalty0 17981--17993, 2021.

\bibitem[Axelrod \& G{\'o}mez-Bombarelli(2022)Axelrod and
  G{\'o}mez-Bombarelli]{geom_drug}
Axelrod, S. and G{\'o}mez-Bombarelli, R.
\newblock Geom, energy-annotated molecular conformations for property
  prediction and molecular generation.
\newblock \emph{Scientific Data}, 9\penalty0 (1):\penalty0 185, 2022.
\newblock \doi{10.1038/s41597-022-01288-4}.
\newblock URL \url{https://doi.org/10.1038/s41597-022-01288-4}.

\bibitem[Dhariwal \& Nichol(2021)Dhariwal and Nichol]{dhariwal2021diffusion}
Dhariwal, P. and Nichol, A.
\newblock Diffusion models beat gans on image synthesis.
\newblock \emph{Advances in Neural Information Processing Systems},
  34:\penalty0 8780--8794, 2021.

\bibitem[Gebauer et~al.(2019)Gebauer, Gastegger, and
  Sch{\"u}tt]{gebauer2019symmetry}
Gebauer, N., Gastegger, M., and Sch{\"u}tt, K.
\newblock Symmetry-adapted generation of 3d point sets for the targeted
  discovery of molecules.
\newblock \emph{Advances in Neural Information Processing Systems}, 32, 2019.

\bibitem[G{\'o}mez-Bombarelli et~al.(2018)G{\'o}mez-Bombarelli, Wei, Duvenaud,
  Hern{\'a}ndez-Lobato, S{\'a}nchez-Lengeling, Sheberla, Aguilera-Iparraguirre,
  Hirzel, Adams, and Aspuru-Guzik]{gomez2018automatic}
G{\'o}mez-Bombarelli, R., Wei, J.~N., Duvenaud, D., Hern{\'a}ndez-Lobato,
  J.~M., S{\'a}nchez-Lengeling, B., Sheberla, D., Aguilera-Iparraguirre, J.,
  Hirzel, T.~D., Adams, R.~P., and Aspuru-Guzik, A.
\newblock Automatic chemical design using a data-driven continuous
  representation of molecules.
\newblock \emph{ACS central science}, 4\penalty0 (2):\penalty0 268--276, 2018.

\bibitem[Guan et~al.(2022)Guan, Qian, Liu, Ma, Ma, and Peng]{guan2021energy}
Guan, J., Qian, W.~W., Liu, Q., Ma, W.-Y., Ma, J., and Peng, J.
\newblock Energy-inspired molecular conformation optimization.
\newblock In \emph{International Conference on Learning Representations}, 2022.

\bibitem[Guan et~al.(2023)Guan, Qian, Peng, Su, Peng, and Ma]{guan2023d}
Guan, J., Qian, W.~W., Peng, X., Su, Y., Peng, J., and Ma, J.
\newblock 3d equivariant diffusion for target-aware molecule generation and
  affinity prediction.
\newblock In \emph{The Eleventh International Conference on Learning
  Representations}, 2023.
\newblock URL \url{https://openreview.net/forum?id=kJqXEPXMsE0}.

\bibitem[Haefeli et~al.(2022)Haefeli, Martinkus, Perraudin, and
  Wattenhofer]{haefeli2022diffusion}
Haefeli, K.~K., Martinkus, K., Perraudin, N., and Wattenhofer, R.
\newblock Diffusion models for graphs benefit from discrete state spaces.
\newblock \emph{arXiv preprint arXiv:2210.01549}, 2022.

\bibitem[Ho et~al.(2020)Ho, Jain, and Abbeel]{ho2020denoising}
Ho, J., Jain, A., and Abbeel, P.
\newblock Denoising diffusion probabilistic models.
\newblock \emph{Advances in Neural Information Processing Systems},
  33:\penalty0 6840--6851, 2020.

\bibitem[Hoogeboom et~al.(2021)Hoogeboom, Nielsen, Jaini, Forr{\'e}, and
  Welling]{hoogeboom2021argmax}
Hoogeboom, E., Nielsen, D., Jaini, P., Forr{\'e}, P., and Welling, M.
\newblock Argmax flows and multinomial diffusion: Learning categorical
  distributions.
\newblock \emph{Advances in Neural Information Processing Systems}, 34, 2021.

\bibitem[Hoogeboom et~al.(2022)Hoogeboom, Satorras, Vignac, and
  Welling]{equivariant_diffusion}
Hoogeboom, E., Satorras, V.~G., Vignac, C., and Welling, M.
\newblock Equivariant diffusion for molecule generation in 3d.
\newblock \emph{arXiv preprint arXiv:2203.17003}, 2022.

\bibitem[Igashov et~al.(2022)Igashov, St{\"a}rk, Vignac, Satorras, Frossard,
  Welling, Bronstein, and Correia]{igashov2022equivariant}
Igashov, I., St{\"a}rk, H., Vignac, C., Satorras, V.~G., Frossard, P., Welling,
  M., Bronstein, M., and Correia, B.
\newblock Equivariant 3d-conditional diffusion models for molecular linker
  design.
\newblock \emph{arXiv preprint arXiv:2210.05274}, 2022.

\bibitem[Jin et~al.(2018)Jin, Barzilay, and Jaakkola]{jin2018junction}
Jin, W., Barzilay, R., and Jaakkola, T.
\newblock Junction tree variational autoencoder for molecular graph generation.
\newblock In \emph{International Conference on Machine Learning}, pp.\
  2323--2332. PMLR, 2018.

\bibitem[Jin et~al.(2020)Jin, Barzilay, and Jaakkola]{jin2020composing}
Jin, W., Barzilay, R., and Jaakkola, T.
\newblock Composing molecules with multiple property constraints.
\newblock \emph{arXiv preprint arXiv:2002.03244}, 2020.

\bibitem[Jo et~al.(2022)Jo, Lee, and Hwang]{jo2022score}
Jo, J., Lee, S., and Hwang, S.~J.
\newblock Score-based generative modeling of graphs via the system of
  stochastic differential equations.
\newblock \emph{arXiv preprint arXiv:2202.02514}, 2022.

\bibitem[Kusner et~al.(2017)Kusner, Paige, and
  Hern{\'a}ndez-Lobato]{kusner2017grammar}
Kusner, M.~J., Paige, B., and Hern{\'a}ndez-Lobato, J.~M.
\newblock Grammar variational autoencoder.
\newblock In \emph{International Conference on Machine Learning}, pp.\
  1945--1954. PMLR, 2017.

\bibitem[Li et~al.(2021)Li, Pei, and Lai]{li2021structure}
Li, Y., Pei, J., and Lai, L.
\newblock Structure-based de novo drug design using 3d deep generative models.
\newblock \emph{Chemical science}, 12\penalty0 (41):\penalty0 13664--13675,
  2021.

\bibitem[Lin et~al.(2022)Lin, Huang, Liu, Li, Ji, and Li]{lin2022diffbp}
Lin, H., Huang, Y., Liu, M., Li, X., Ji, S., and Li, S.~Z.
\newblock Diffbp: Generative diffusion of 3d molecules for target protein
  binding.
\newblock \emph{arXiv preprint arXiv:2211.11214}, 2022.

\bibitem[Liu et~al.(2022)Liu, Luo, Uchino, Maruhashi, and Ji]{liu2022graphbp}
Liu, M., Luo, Y., Uchino, K., Maruhashi, K., and Ji, S.
\newblock Generating 3d molecules for target protein binding.
\newblock In \emph{International Conference on Machine Learning}, 2022.

\bibitem[Liu et~al.(2018)Liu, Allamanis, Brockschmidt, and
  Gaunt]{liu2018constrained}
Liu, Q., Allamanis, M., Brockschmidt, M., and Gaunt, A.
\newblock Constrained graph variational autoencoders for molecule design.
\newblock \emph{Advances in neural information processing systems}, 31, 2018.

\bibitem[Luo et~al.(2021)Luo, Guan, Ma, and Peng]{luo20213d}
Luo, S., Guan, J., Ma, J., and Peng, J.
\newblock A 3d generative model for structure-based drug design.
\newblock \emph{Advances in Neural Information Processing Systems}, 34, 2021.

\bibitem[Luo \& Ji(2022)Luo and Ji]{luo2022autoregressive}
Luo, Y. and Ji, S.
\newblock An autoregressive flow model for 3d molecular geometry generation
  from scratch.
\newblock In \emph{International Conference on Learning Representations
  (ICLR)}, 2022.

\bibitem[Malinin \& Gales(2018)Malinin and Gales]{uncertainty}
Malinin, A. and Gales, M.
\newblock Predictive uncertainty estimation via prior networks.
\newblock \emph{Advances in neural information processing systems}, 31, 2018.

\bibitem[Nichol \& Dhariwal(2021)Nichol and Dhariwal]{nichol2021improved}
Nichol, A.~Q. and Dhariwal, P.
\newblock Improved denoising diffusion probabilistic models.
\newblock In \emph{International Conference on Machine Learning}, pp.\
  8162--8171. PMLR, 2021.

\bibitem[Niu et~al.(2020)Niu, Song, Song, Zhao, Grover, and
  Ermon]{niu2020permutation}
Niu, C., Song, Y., Song, J., Zhao, S., Grover, A., and Ermon, S.
\newblock Permutation invariant graph generation via score-based generative
  modeling.
\newblock In \emph{International Conference on Artificial Intelligence and
  Statistics}, pp.\  4474--4484. PMLR, 2020.

\bibitem[O'Boyle et~al.(2011)O'Boyle, Banck, James, Morley, Vandermeersch, and
  Hutchison]{o2011obabel}
O'Boyle, N.~M., Banck, M., James, C.~A., Morley, C., Vandermeersch, T., and
  Hutchison, G.~R.
\newblock Open babel: An open chemical toolbox.
\newblock \emph{Journal of cheminformatics}, 3\penalty0 (1):\penalty0 1--14,
  2011.

\bibitem[Peng et~al.(2022)Peng, Luo, Guan, Xie, Peng, and
  Ma]{peng2022pocket2mol}
Peng, X., Luo, S., Guan, J., Xie, Q., Peng, J., and Ma, J.
\newblock {P}ocket2{M}ol: Efficient molecular sampling based on 3{D} protein
  pockets.
\newblock In Chaudhuri, K., Jegelka, S., Song, L., Szepesvari, C., Niu, G., and
  Sabato, S. (eds.), \emph{Proceedings of the 39th International Conference on
  Machine Learning}, volume 162 of \emph{Proceedings of Machine Learning
  Research}, pp.\  17644--17655. PMLR, 17--23 Jul 2022.
\newblock URL \url{https://proceedings.mlr.press/v162/peng22b.html}.

\bibitem[Ragoza et~al.(2020)Ragoza, Masuda, and Koes]{ragoza2020learning}
Ragoza, M., Masuda, T., and Koes, D.~R.
\newblock Learning a continuous representation of 3d molecular structures with
  deep generative models.
\newblock \emph{arXiv preprint arXiv:2010.08687}, 2020.

\bibitem[Ramesh et~al.(2022)Ramesh, Dhariwal, Nichol, Chu, and
  Chen]{ramesh2022hierarchical}
Ramesh, A., Dhariwal, P., Nichol, A., Chu, C., and Chen, M.
\newblock Hierarchical text-conditional image generation with clip latents.
\newblock \emph{arXiv preprint arXiv:2204.06125}, 2022.

\bibitem[Roney et~al.(2022)Roney, Maragakis, Skopp, and
  Shaw]{roney2022generating}
Roney, J.~P., Maragakis, P., Skopp, P., and Shaw, D.~E.
\newblock Generating realistic 3d molecules with an equivariant conditional
  likelihood model, 2022.
\newblock URL \url{https://openreview.net/forum?id=Snqhqz4LdK}.

\bibitem[Satorras et~al.(2021{\natexlab{a}})Satorras, Hoogeboom, Fuchs, Posner,
  and Welling]{satorras2021enf}
Satorras, V.~G., Hoogeboom, E., Fuchs, F.~B., Posner, I., and Welling, M.
\newblock E (n) equivariant normalizing flows.
\newblock \emph{arXiv preprint arXiv:2105.09016}, 2021{\natexlab{a}}.

\bibitem[Satorras et~al.(2021{\natexlab{b}})Satorras, Hoogeboom, and
  Welling]{satorras2021egnn}
Satorras, V.~G., Hoogeboom, E., and Welling, M.
\newblock E (n) equivariant graph neural networks.
\newblock In \emph{International Conference on Machine Learning}, pp.\
  9323--9332. PMLR, 2021{\natexlab{b}}.

\bibitem[Schneuing et~al.(2022)Schneuing, Du, Harris, Jamasb, Igashov, Du,
  Blundell, Li{\'o}, Gomes, Welling, et~al.]{schneuing2022structure}
Schneuing, A., Du, Y., Harris, C., Jamasb, A., Igashov, I., Du, W., Blundell,
  T., Li{\'o}, P., Gomes, C., Welling, M., et~al.
\newblock Structure-based drug design with equivariant diffusion models.
\newblock \emph{arXiv preprint arXiv:2210.13695}, 2022.

\bibitem[Segler et~al.(2018)Segler, Kogej, Tyrchan, and
  Waller]{segler2018generating}
Segler, M.~H., Kogej, T., Tyrchan, C., and Waller, M.~P.
\newblock Generating focused molecule libraries for drug discovery with
  recurrent neural networks.
\newblock \emph{ACS central science}, 4\penalty0 (1):\penalty0 120--131, 2018.

\bibitem[Shi et~al.(2020)Shi, Xu, Zhu, Zhang, Zhang, and Tang]{shi2020graphaf}
Shi, C., Xu, M., Zhu, Z., Zhang, W., Zhang, M., and Tang, J.
\newblock Graphaf: a flow-based autoregressive model for molecular graph
  generation.
\newblock \emph{arXiv preprint arXiv:2001.09382}, 2020.

\bibitem[Song \& Ermon(2019)Song and Ermon]{song2019generative}
Song, Y. and Ermon, S.
\newblock Generative modeling by estimating gradients of the data distribution.
\newblock \emph{Advances in Neural Information Processing Systems}, 32, 2019.

\bibitem[Song et~al.(2020)Song, Sohl-Dickstein, Kingma, Kumar, Ermon, and
  Poole]{song2020score}
Song, Y., Sohl-Dickstein, J., Kingma, D.~P., Kumar, A., Ermon, S., and Poole,
  B.
\newblock Score-based generative modeling through stochastic differential
  equations.
\newblock \emph{arXiv preprint arXiv:2011.13456}, 2020.

\bibitem[Vignac et~al.(2022)Vignac, Krawczuk, Siraudin, Wang, Cevher, and
  Frossard]{vignac2022digress}
Vignac, C., Krawczuk, I., Siraudin, A., Wang, B., Cevher, V., and Frossard, P.
\newblock Digress: Discrete denoising diffusion for graph generation.
\newblock \emph{arXiv preprint arXiv:2209.14734}, 2022.

\bibitem[Vignac et~al.(2023)Vignac, Osman, Toni, and Frossard]{vignac2023midi}
Vignac, C., Osman, N., Toni, L., and Frossard, P.
\newblock Midi: Mixed graph and 3d denoising diffusion for molecule generation.
\newblock \emph{arXiv preprint arXiv:2302.09048}, 2023.

\bibitem[Weininger(1988)]{weininger1988smiles}
Weininger, D.
\newblock Smiles, a chemical language and information system. 1. introduction
  to methodology and encoding rules.
\newblock \emph{Journal of chemical information and computer sciences},
  28\penalty0 (1):\penalty0 31--36, 1988.

\bibitem[Wu et~al.(2022)Wu, Gong, Liu, Ye, and qiang liu]{moldiffbridge}
Wu, L., Gong, C., Liu, X., Ye, M., and qiang liu.
\newblock Diffusion-based molecule generation with informative prior bridges.
\newblock In Oh, A.~H., Agarwal, A., Belgrave, D., and Cho, K. (eds.),
  \emph{Advances in Neural Information Processing Systems}, 2022.
\newblock URL \url{https://openreview.net/forum?id=TJUNtiZiTKE}.

\bibitem[Xu et~al.(2022)Xu, Yu, Song, Shi, Ermon, and Tang]{xu2022geodiff}
Xu, M., Yu, L., Song, Y., Shi, C., Ermon, S., and Tang, J.
\newblock Geodiff: A geometric diffusion model for molecular conformation
  generation.
\newblock \emph{arXiv preprint arXiv:2203.02923}, 2022.

\bibitem[You et~al.(2018)You, Liu, Ying, Pande, and Leskovec]{you2018graph}
You, J., Liu, B., Ying, R., Pande, V., and Leskovec, J.
\newblock Graph convolutional policy network for goal-directed molecular graph
  generation.
\newblock \emph{arXiv preprint arXiv:1806.02473}, 2018.

\bibitem[Zhou et~al.(2019)Zhou, Kearnes, Li, Zare, and
  Riley]{zhou2019optimization}
Zhou, Z., Kearnes, S., Li, L., Zare, R.~N., and Riley, P.
\newblock Optimization of molecules via deep reinforcement learning.
\newblock \emph{Scientific reports}, 9\penalty0 (1):\penalty0 1--10, 2019.

\end{thebibliography}
\bibliographystyle{icml2023}


\newpage
\appendix
\onecolumn

\section{More Details about the Model} \label{sec: app_model}

\subsection{Hyper-parameters of the Neural Networks and Training} \label{sec: app_hyper}

We set the embedding dimensions of node types and edge types as 256 and 64, respectively and all intermediate hidden dimensions are constant. The time embedding dimensions are 10. The graph neural networks contain six layers. We trained the diffusion network using AdamW optimizer with a learning rate $1\times 10^{-4}$ and batch size 256 for $110,000$ iterations. For the weights of the atom loss and bond loss, i.e., $\lambda_1$ and $\lambda_2$, we set $\lambda_1=\lambda_2=100$ so that the losses of atom types, atom positions, and bond types were almost in the same magnitude. The source codes will be provided at \url{https://github.com/pengxingang/MolDiff}.

\subsection{Details about the Diffusion Process} \label{sec: app_diff}

By applying Bayes' theorem, we derive the posterior $q(M^{t-1}|M^{t-1}, M^{0})$ from Eq. \ref{eq: diff_one_step} and Eq. \ref{eq: diff_t} as:
\begin{equation}
\begin{aligned}
    q_({a}_i^{t-1} | {a}_i^{t}, {a}_i^{0}) &= \mathcal{C}( {a}_i^{t-1} ; \Theta({a}_i^{t}, {a}_i^{0})) \\
    q_(\mathbf{r}_i^{t-1} | \mathbf{r}_i^{t}, \mathbf{r}_i^{0}) &= \mathcal{N}( \mathbf{r}_i^{t-1} | \boldsymbol{\tilde{\mu}}^{t} (\mathbf{r}_i^{0}, \mathbf{r}_i^{t}), \tilde{\beta}^{t}) \\
    q_({b}_{ij}^{t-1} | {b}_{ij}^{t}, {b}_{ij}^{0}),  &= \mathcal{C}( {b}_{ij}^{t-1} ; \Theta({b}_{ij}^{t}, {b}_{ij}^{0}) ),
\end{aligned}
\end{equation}
where 
\begin{equation}
\begin{aligned}
    \tilde{\boldsymbol{\mu}}^{t} (\mathbf{r}_i^{0}, \mathbf{r}_i^{t}) :&= \frac{\sqrt{\bar{\alpha}^{t-1}\beta^{t}}}{1 - \bar{\alpha}^{t}} \mathbf{r}_i^{0} + \frac{\sqrt{\alpha^{t}}(1 - \bar{\alpha}^{t-1})}{1 - \bar{\alpha}^{t}} \mathbf{r}_i^{t} \\
    \tilde{\beta}^{t} :&= \frac{1 - \bar{\alpha}^{t-1}}{1 - \bar{\alpha}^{t}}\beta^{t} \\
    \Theta(\mathbf{a}_{i}^{t}, \mathbf{a}_{i}^{0}) &\propto \left[ \alpha^{t} \mathbf{a}_{i}^{t} + (1 - \alpha^{t}) \mathbbm{1}_{k}) \right] \cdot \left[ \bar{\alpha}^{t-1} \mathbf{a}_{i}^{t} + ( 1 - \bar{\alpha}^{t-1}) \mathbbm{1}_{k} ) \right] \\
    \Theta(\mathbf{b}_{ij}^{t}, \mathbf{b}_{ij}^{0}) &\propto \left[ \alpha^{t} \mathbf{b}_{ij}^{t} + (1 - \alpha^{t}) \mathbbm{1}_{k'}) \right] \cdot \left[ \bar{\alpha}^{t-1} \mathbf{b}_{ij}^{t} + ( 1 - \bar{\alpha}^{t-1}) \mathbbm{1}_{k'} ) \right],
\end{aligned}
\end{equation}
and the symbol ``$\cdot$'' means element-wise multiplication of two vectors.

\subsection{Details about the Model Architectures} \label{sec: app_model}

Here we detailedly describe the architectures of the neural networks defined in Eq. \ref{eq: model}, i.e., $\phi_d$, $\phi_v$, $\phi_e$, and $\phi_r$.

The network $\phi_d$ takes as input the edge feature vector $\mathbf{e}_{ij}$ and the edge lengths $\| \mathbf{r}_i - \mathbf{r}_j \|$. It applies RBF kernels to the distance and then combines them with the input feature vector. Finally, it applies a linear layer to process the concatenated feature to the new feature vector $\tilde{\mathbf{e}}_{ij}$.

The network $\phi_v$ takes as input the vertex feature vectors $\mathbf{v}_j$, the edge feature vector $\tilde{\mathbf{e}}_{ij}$, and the step $t$ and outputs the message from vertex $j$ to vertex $i$ (denoted as $\mathbf{m}_{ij}$):
\begin{equation}
\begin{aligned}
    \mathbf{v}_j' &= \text{MLP}(\mathbf{v}_j) \\
    \tilde{\mathbf{e}}_{ij}' &= \text{MLP}(\tilde{\mathbf{e}}_{ij}) \\
    \mathbf{g}_{ij} &= \text{MLP}(\text{concat}(\mathbf{v}_j, \tilde{\mathbf{e}}_{ij}, t))  \\
    \mathbf{m}_{ij} &= \text{Linear}(\mathbf{v}_j' \cdot \tilde{\mathbf{e}}_{ij}') \cdot \text{sigmoid}(\mathbf{g}_{ij}),
\end{aligned}
\end{equation}
where $\text{concat}(\cdot)$ means concatenating along the feature dimension and $\text{sigmoid}(\cdot)$ is the sigmoid activation function. 

The network $\phi_e$ takes as input the vertex feature vectors $\mathbf{v}_i$, the edge feature vector $\tilde{\mathbf{e}}_{ij}$, and the step $t$ and outputs the message from the edge $ij$ (denoted as $\mathbf{m}_{ij}$):
\begin{equation}
\begin{aligned}
    \mathbf{m}_{ij}' &= \text{MLP}( \text{Linear}(\mathbf{v}_i) \cdot \text{Linear}(\tilde{\mathbf{e}}_{ij}) ) \\
    \mathbf{g}_{ij} &= \text{MLP}(\text{concat}(\mathbf{v}_i, \tilde{\mathbf{e}}_{ij}, t)) \\
    \mathbf{m}_{ij} &= \mathbf{m}_{ij}' \cdot \mathbf{g}_{ij}.
\end{aligned}
\end{equation}

Finally, the network $\phi_r$ takes as input the vertex feature vectors $\mathbf{v}_i, \mathbf{v}_j$, the edge feature vector $\tilde{\mathbf{e}}_{ij}$ and step $t$ and outputs a scalar $m_{ij}$:
\begin{equation}
\begin{aligned}
    \mathbf{v}_i' &= \text{MLP}(\mathbf{v}_i) \\
    \mathbf{v}_j' &= \text{MLP}(\mathbf{v}_j) \\
    m_{ij} &= \phi_e(\mathbf{v}_i' \cdot \mathbf{v}_j', \tilde{\mathbf{e}}_{ij}, t)
\end{aligned}
\end{equation}

\subsection{Details about the Noise Schedule} \label{sec: app_noise}

\begin{figure}[htbp]
    \centering
    \includegraphics[width=0.4\linewidth]{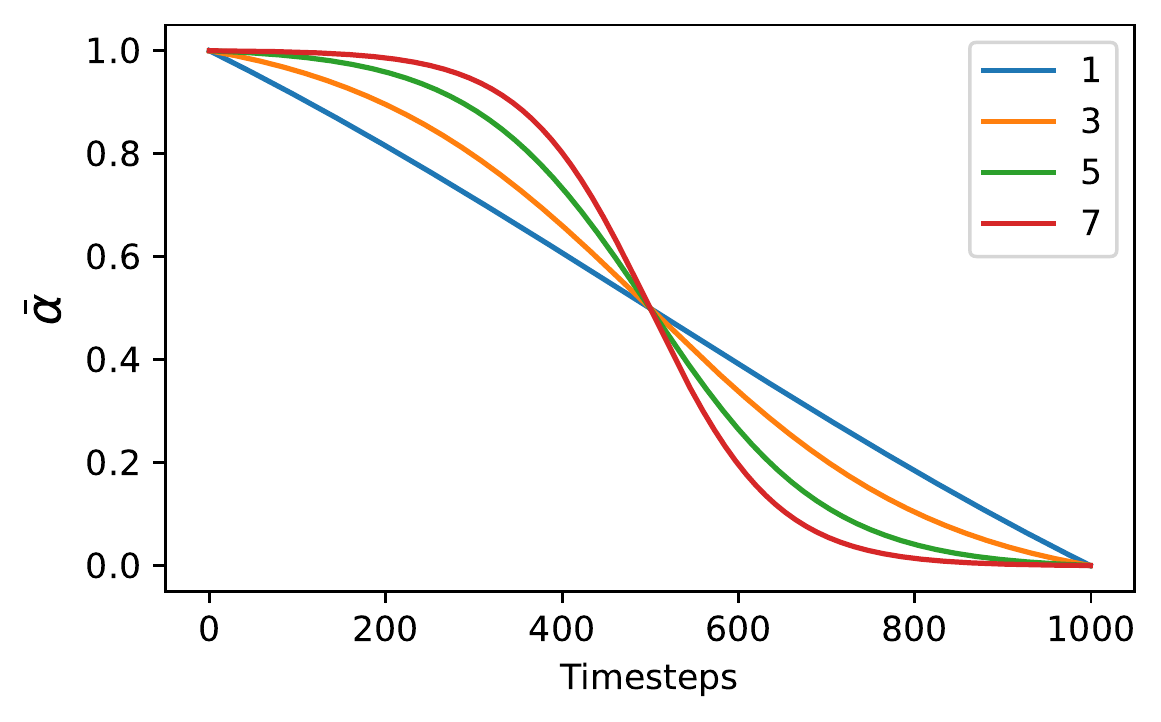}
    \caption{The noise level $\bar{\alpha}$ curves with different parameter $w$}
    \label{fig: schedule}
\end{figure}

The cosine noise schedule is a popular schedule that defines the schedule of $\bar{\alpha}^{t}$ using a cosine function and achieves better performance than other schedules \cite{nichol2021improved}. However, the cosine schedule only has one tuneable hyper-parameter and thus it is hard to adapt the shape and the range of the schedule curve of $\bar{\alpha}^{t}$. We define a new schedule scheme that utilizes the sigmoid function and possess three tuneable hyper-parameters $s_1, s_T, w$ to adapt the schedule curve. The schedule is defined as:
\begin{equation}
\begin{aligned}
    s &= (s_T - s_1) / (\text{sigmoid}(-w) - \text{sigmoid}(w)) \\
    b &= 0.5 \times ( s_1 + s_T - s ) \\
    \bar{\alpha}_t &= s \times \text{sigmoid}(-w (2t/T-1)) + b.
\end{aligned}
\end{equation}
The parameters $s_1$ and $s_T$ represent the values of $\bar{\alpha}_1$ and $\bar{\alpha}_T$, respectively, controlling the range of the noise. The parameter $w$ influences the slope at medium steps and controls how fast $\bar{\alpha}_t$ changes at the beginning and the end of the diffusion process. We show how the parameter $w$ influences the shapes in Fig. \ref{fig: schedule}.

\begin{figure}[htb]
    \centering
    \includegraphics[width=0.4\linewidth]{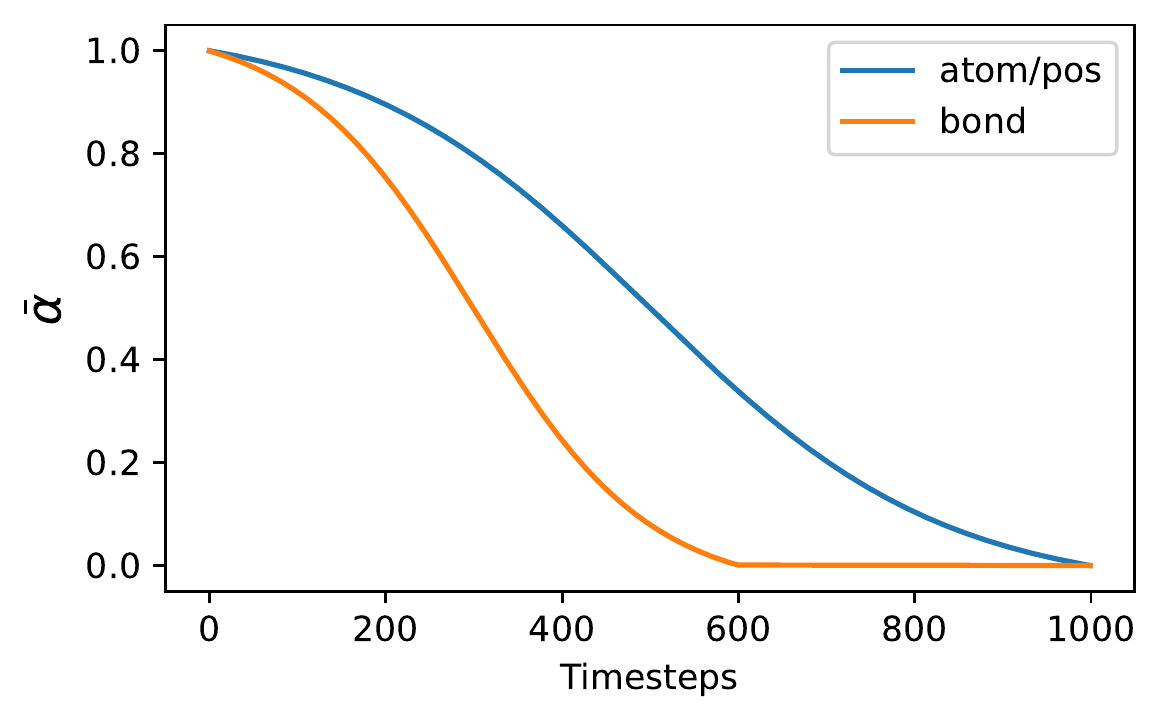}
    \caption{The noise level $\bar{\alpha}$ curves used by \model{}}
    \label{fig: sche_real}
\end{figure}

In our implementation, we chose the parameters $s_1=0.9999,s_T=0.0001,w=3$ for atom types and atom positions for the whole diffusion process $t\in [1, T]$. For the bond type, we used $s_1=0.9999,s_T=0.001,w=3$ during diffusion steps $[1, 600]$ in the first stage and $s_1=0.001,s_T=0.0001,w=2$ during steps $[600, 1000]$ in the second stage. The curves of \model{} are shown in Fig. \ref{fig: sche_real}

\subsection{Details about the Bond Predictor} \label{sec: app_bondpred}

The bond predictor has almost the same architecture as the main diffusion neural networks except for two differences. First, the bond predictor does not need the position updates. Second, the initial input edge features do not contain the bond types of the edge. Instead, we provide the atom types of the two ends of the edges as the initial edge features.

The bond predictor was trained using the same dataset as the diffusion network. The training uses the same noise schedules of the atom positions and atom types. At step $t$, the bond predictor takes as input the noised atom positions $R_t$, atom types $A_t$ and step $t$ and predicts the bond types at step 0, i.e., $\hat{B}_0$. The loss function is the cross-entropy between the predicted bond types and the ground truth.  We trained the bond predictor using AdamW optimizer with a learning rate $1\times 10^{-4}$ and batch size 256 for $300,000$ iterations.

\section{Preprocessing of the Geom-Drug Dataset} \label{sec: app_data}

We downloaded the GEOM-Drug from the database website \cite{geom_drug}. We filtered the molecules through the following criterion: (1) can be parsed by the RDKit package; (2) not broken; (3) the number of heavy atoms within $[8, 60]$; (4) not contain the elements other than H, C, N, O, F, P, S, and Cl; (5) not contain chemical bonds other than single, double, triple and aromatic bonds. After filtering, we removed the hydrogen atoms and constructed the training, validation, and testing datasets with 231523, 28941, and 28940 molecules, respectively.

\section{More Details and Results about the Experiments} \label{sec: app_exper}

\subsection{Details about the Experiment Settings and Baselines} \label{sec: app_baseline}

In all experiments, we kept generating molecules until there were 1000 valid and complete molecules and then calculated metrics according to their definitions. We repeated this for three times and reported the average values in the main text and standard deviations in the Appendix \ref{sec: app_std}.

We re-trained the baseline EDM using our splittings of the GEOM-Dataset dataset. For the variants of \model{} that did not incorporate the bond types during diffusion, we removed the edge updates in Eq. \ref{eq: model}, and the initial edge embedding did not contain the information of bond type. Three different bond post-predicting strategies were considered for three different models: First, a lookup table that was the same as the one used by EDM was utilized to match the bond lengths with the atom distances. Second, the NN-based bond predictor that was originally used for the guidance was applied to predict the bond types after diffusion. Third, the chemical toolkit Open Babel was applied to add the chemical bonds. For the variants of \model{} that used continuous diffusion for both atom and bond types, they regarded the one-hot embedding of the types as continuous values in the latent spaces and added Gaussian noise. The scaling for continuous diffusion was to divide the one-hot embedding of the atom and bond types by scalars before diffusion, which was proposed by the authors of EDM. We set the scaling for the atom types, atom positions, and bond types as 1, 4, and 8, respectively. For the variant with bond length loss, we further added an MSE loss to minimize the difference between the real bond lengths $\|\mathbf{r}_i^{t-1} - \mathbf{r}_j^{t-1}\|_2$ with the predicted ones $\|\boldsymbol{\mu}_\theta(M^t, t)_i - \boldsymbol{\mu}_\theta(M^t, t)_j\|_2$ for $ij$ pairs that $b_{ij}^{0}$ were not none-type, which is a simple method to attempt to generate more accurate bond lengths. Note all these variants did not use the newly proposed noise schedule and the guidance.

\subsection{Details about the Metrics} \label{sec: app_metric}

For the metric of drug-likeness, we borrowed the evaluation code from Pocket2Mol \cite{peng2022pocket2mol}. For the analyses of bond lengths, the SMARTS (SMILES arbitrary target specification) of considered bond types were: c:c, [\#6]-[\#6], [\#6]-[\#7], [\#6]-O, c:n, [\#6]=O, [\#6]-S, O=S, c:o, c:s, [\#6]-F, n:n, [\#6]-Cl, [\#6]=[\#6], [\#7]-S, [\#6]=[\#7], [\#7]-[\#7], [\#7]-O, [\#6]=S, and [\#7]=O. For the analyses of bond angles, the SMARTS of considered bond pairs were: c:c:c, [\#6]-[\#6]-[\#6], [\#6]-[\#7]-[\#6], [\#7]-[\#6]-[\#6], c:c-[\#6], [\#6]-O-[\#6], O=[\#6]-[\#6], [\#7]-c:c, n:c:c, c:c-O, c:n:c, [\#6]-[\#6]-O, and O=[\#6]-[\#7]. The SMARTS of considered bond triplets were: c:c:c:c, [\#6]-[\#6]-[\#6]-[\#6], [\#6]-[\#7]-[\#6]-[\#6], [\#6]-c:c:c, [\#7]-[\#6]-[\#6]-[\#6], [\#7]-c:c:c, O-c:c:c, [\#6]-[\#7]-c:c, [\#7]-[\#6]-c:c, n:c:c:c, [\#6]-[\#7]-[\#6]=O, [\#6]-[\#6]-c:c, c:c-[\#7]-[\#6], c:n:c:c, and [\#6]-O-c:c.

\subsection{Details and More Results of the Ablation Study} \label{sec: app_ablation}

For the variant without guidance, we removed the guidance term in Eq. \ref{eq: gui} during generation. For the variant without the new noise schedule, the schedule $\bar{\alpha}_t$ of bond type was the same as that of the atoms. In Tab. \ref{tab: ablation_add}, we reported the generation abilities and the drug-likeness properties of the ablation study, which indicated that the our proposed noise schedule and the guidance of the bond predictor did not damage these basic properties.

\begin{table}[!ht]
    \caption{More metrics of the ablation study.}
    \centering
    \begin{tabular}{l|p{1cm}p{1cm}p{1cm}|p{1cm}p{1cm}p{1cm}p{1cm}|p{1cm}p{1cm}p{1cm}}
    \toprule
        ~ & Validity & Connec-tivity & Succ. rate & Novelty & Diversity & Unique- ness & Sim. Val. & QED & SA & Lipinski  \\ \hline
        MolDiff & 0.997  & 0.996  & 0.993  & 0.972  & 0.769  & 0.986  & 0.634  & 0.668  & 0.874  & 4.986   \\ 
        No Gui. & 0.992  & 0.964  & 0.956  & 0.983  & 0.763  & 0.995  & 0.603  & 0.680  & 0.825  & 4.985   \\ 
        No Sche. & 1.000  & 0.993  & 0.993  & 0.985  & 0.766  & 0.998  & 0.639  & 0.692  & 0.862  & 4.951   \\ 
        Neither & 0.999  & 0.971  & 0.970  & 0.990  & 0.764  & 0.999  & 0.628  & 0.710  & 0.822  & 4.941  \\
    \bottomrule
    \end{tabular}
    \label{tab: ablation_add}
\end{table}

\subsection{Standard Deviations of the Values in the Main Text.} \label{sec: app_std}

To facilitate the comparison of the metric values of different models, here we provided the standard deviations of those values in the tables of the Sec. \ref{sec: results} (i.e., Tab. \ref{tab: gen_ability}, \ref{tab: drug_prop}, and \ref{tab: ablation}) in Tab. \ref{tab: gen_ability_std}, \ref{tab: drug_prop_std}, and \ref{tab: ablation_std}.

\begin{table}[!ht]
    \centering
    \caption{The standard deviations of the generation abilities of the models.}
    \begin{tabular}{l|cc|cccc} 
    \toprule
        ~ & Validity & Connectivity & Novelty & Diversity \\ \hline
        EDM & 0.0002  & 0.0068  & 0.0000  & 0.0033  \\
        MolDiff & 0.0012  & 0.0014  & 0.0024  & 0.0004  \\ \hline
        Predict bond(Lookup table) & 0.0035  & 0.0075  & 0.0000  & 0.0045  \\ 
        Predict bond(NN Predictor) & 0.0016  & 0.0027  & 0.0065  & 0.0025  \\ 
        Predict bond(Openbabel) & 0.0023  & 0.0043  & 0.0037  & 0.0024  \\ 
        Conti. Diffusion & 0.0116  & 0.0142  & 0.0004  & 0.0010  \\ 
        Conti. Diffusion(scaling) & 0.0014  & 0.0027  & 0.0005  & 0.0039  \\ 
        Add bond length loss & 0.0008  & 0.0030  & 0.0021  & 0.0028  \\ 
    \bottomrule
    \end{tabular}
    \label{tab: gen_ability_std}
\end{table}
\begin{table}[!ht]
    \centering
    \caption{The standard deviations of the qualities of the generated molecules}
    \begin{tabular}{l|l|ll}
    \toprule
        Group & Metrics & EDM & MolDiff  \\ \hline
        \multirow{3}*{\makecell[l]{Drug-likeness}} & QED & 0.0059  & 0.0050   \\ 
        ~ & SA & 0.0038  & 0.0021   \\ 
        ~ & Lipinski & 0.0076  & 0.0029   \\ \hline
        \multirow{4}*{\makecell[l]{3D structures}} & RMSD & 0.0121  & 0.0156   \\ 
        ~ & JS. bond lengths & 0.0033  & 0.0079   \\ 
        ~ & JS. bond angels & 0.0120  & 0.0070   \\ 
        ~ & JS. dihedral angles & 0.0031  & 0.0027   \\ \hline
        \multirow{5}*{\makecell[l]{Bonds}} & JS. \#bonds per atoms & 0.0187  & 0.0051   \\ 
        ~ & JS. basic bond types & 0.0016  & 0.0027   \\ 
        ~ & JS. freq. bond types & 0.0012  & 0.0019   \\ 
        ~ & JS. freq. bond angles & 0.0028  & 0.0025   \\ 
        ~ & JS. freq. bond triplets & 0.0043  & 0.0025   \\ \hline
        \multirow{3}*{\makecell[l]{Rings}} & JS. \#rings & 0.0179  & 0.0040   \\ 
        ~ & JS. \#n-sized rings & 0.0034  & 0.0020   \\ 
        ~ & \#Intersecting rings & 0.4714  & 0.0000  \\
    \bottomrule
    \end{tabular}
    \label{tab: drug_prop_std}
\end{table}

\begin{table}[!ht]
    \caption{The standard deviations of the values in the ablation study}
    \centering
    \begin{tabular}{llll||llll}
    \toprule
        Metrics & No Gui. & No Sche. & Neither & Metrics & No Gui. & No Sche. & Neither  \\ \hline
        Validity & 0.0009  & 0.0005  & 0.0000  & JS. bond lengths & 0.0152  & 0.0066  & 0.0133   \\
        Connectivity & 0.0025  & 0.0008  & 0.0049  & JS. bond angels & 0.0052  & 0.0060  & 0.0057   \\
        Novelty & 0.0012  & 0.0038  & 0.0005  & JS. dihedral angles & 0.0034  & 0.0041  & 0.0043   \\
        Diversity & 0.0012  & 0.0023  & 0.0014  & JS. \#bonds per atoms & 0.0008  & 0.0051  & 0.0108   \\
        Uniqueness & 0.0016  & 0.0017  & 0.0000  & JS. basic bond types & 0.0026  & 0.0019  & 0.0017   \\ 
        Sim. Val & 0.0023  & 0.0018  & 0.0012  & JS. freq. bond types & 0.0023  & 0.0038  & 0.0024   \\ 
        QED & 0.0024  & 0.0016  & 0.0052  & JS. freq. bond angles & 0.0026  & 0.0031  & 0.0032   \\ 
        SA & 0.0040  & 0.0004  & 0.0030  & JS. freq. bond triplets & 0.0041  & 0.0029  & 0.0033   \\ 
        Lipinski & 0.0016  & 0.0028  & 0.0048  & JS. \#rings & 0.0067  & 0.0056  & 0.0032   \\ 
        RMSD & 0.0174  & 0.0056  & 0.0093  & JS. \#n-sized rings & 0.0033  & 0.0017  & 0.0048   \\ 
        ~ & ~ & ~ & ~ & \#Intersecting rings & 0.4714  & 0.0000  & 0.0000  \\
    \bottomrule
    \end{tabular}
    \label{tab: ablation_std}
\end{table}

\subsection{More Examples of Generated Molecules}

\begin{figure}[htbp]
    \centering
    \includegraphics[width=\linewidth]{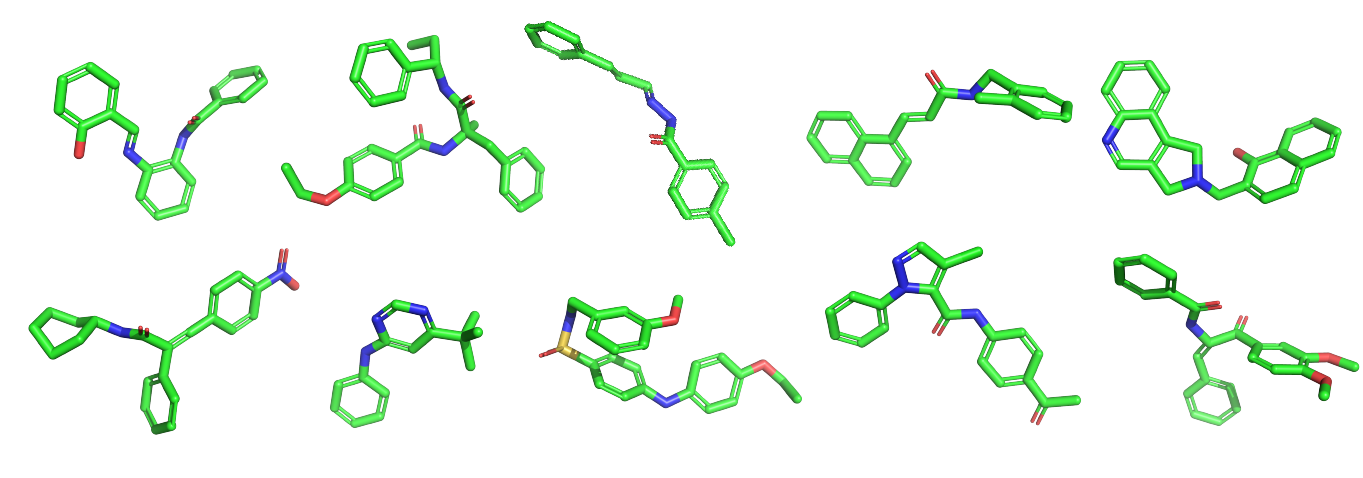}
    \caption{Example of 3D molecules generated by \model{}}
    \label{fig: 3dmol}
\end{figure}

\begin{figure}[htbp]
    \centering
    \includegraphics[width=\linewidth]{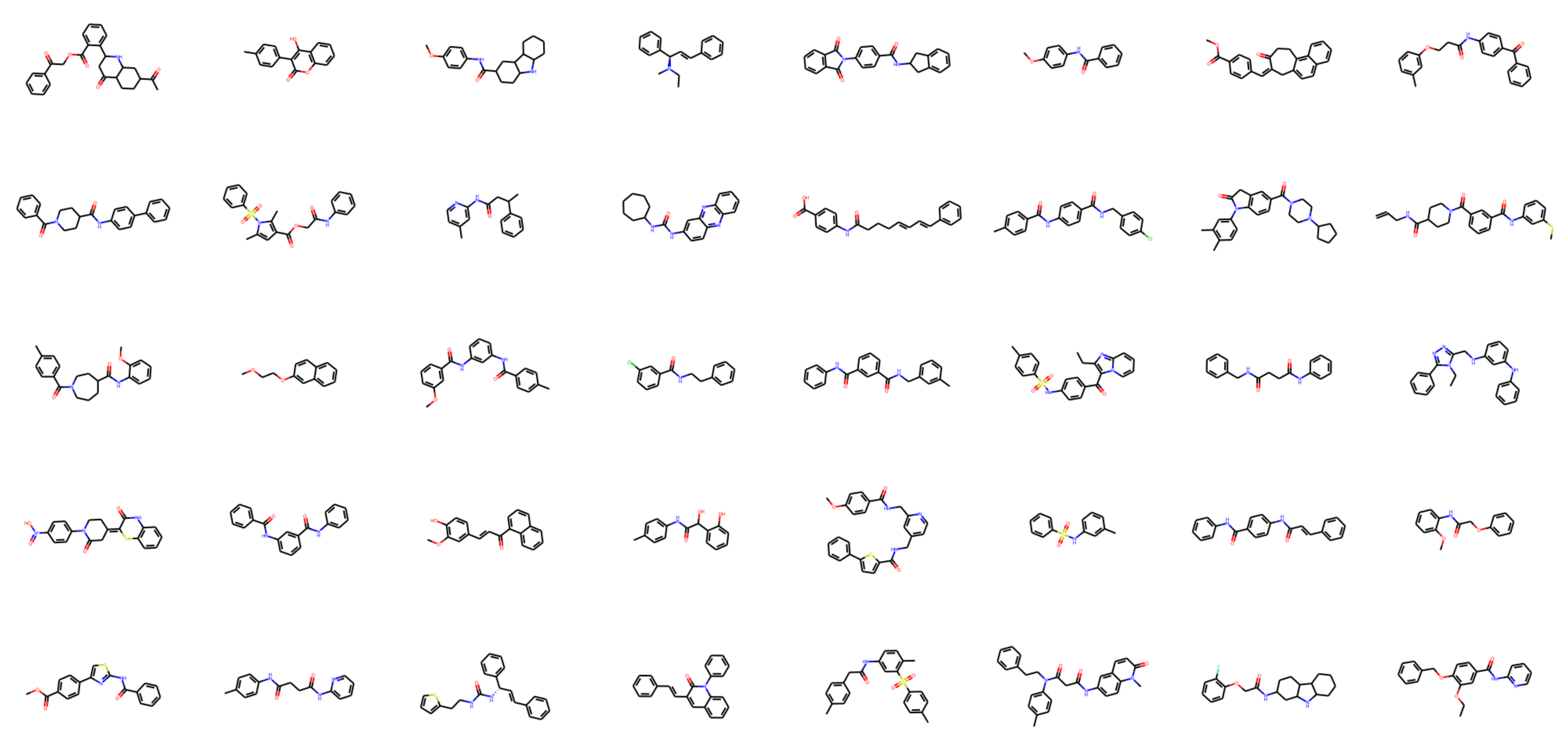}
    \caption{Examples of the graphs of the molecules generated by \model{}}
    \label{fig: mol_gen}
\end{figure}

We showed the 3D conformations and the 2D graphs of several molecules generated by \model{} in Fig. \ref{fig: 3dmol} and Fig. \ref{fig: mol_gen}.

\section{Additional evaluations on more datasets} \label{sec: more_eval}

In the main text, we assessed models on the molecules only containing major elements in the GEOM-Drug dataset and removed hydrogen atoms when training and evaluation. Here, we conducted additional experiments on three more datasets: 1) molecules in the GEOM-Drug dataset containing major elements but with hydrogens reserved; 2) molecules in the GEOM-Drug dataset containing all elements; 3) a different dataset QM9 dataset. Note in these experiments, the sampling of \model{} did not use the guidance of bond predictor.

\subsection{Evaluation on Molecules with Hydrogens Reserved}

The reason that we removed the hydrogens was from a practical perspective because hydrogens can be easily speculated from the heavy atoms and were often omitted in molecule representations in practical scenarios. However, the inclusion of hydrogens resulted in bigger graphs for a more rigid benchmark, which can help us understand the behaviors of models. Therefore, we have added an experiment to incorporate all hydrogen atoms.

As shown in Tab. \ref{tab: with_h}, compared to the performance without hydrogens (the result of one repeat in our original dataset), \model{} performed worse in the dataset with hydrogens, especially the connectivity and thus the success rate. This discrepancy was caused by the fact that the inclusion of hydrogens made the average number of atoms per molecule raised from around 24 to around 45, resulting in much more complicated molecular graphs. But the success rate of \model{} with hydrogens (0.739) was still higher than EDM even without hydrogens (0.371 in Tab. \ref{tab: gen_ability}) and the qualities of the generated molecules were not obviously influenced by the inclusion of hydrogens.

\begin{table}[htb]
    \centering
    \caption{Evaluation of \model{} on molecules with or without hydrogens}
    \begin{tabular}{c|cc}
    \toprule
    Dataset & With hydrogen & Without hydrogen \\\hline
    Validity & 0.993 & 0.957 \\
    Connectivity & 0.961 & 0.772 \\
    Succ. Rate & 0.954 & 0.739 \\
    Novelty & 0.983 & 1.000  \\
    Uniquness & 0.997 & 1.000  \\
    Diversity & 0.761 & 0.427  \\
    Sim. Val. &	0.603 &	0.695  \\ \hline
    QED & 0.677 & 0.688  \\ 
    SA & 0.821 & 0.806  \\
    Lipinski & 4.983 & 4.868  \\ \hline
    RMSD. & 1.005 & 1.032  \\
    JS. bond lengths & 0.374 & 0.414  \\
    JS. bond angles	 & 0.168 & 0.182  \\
    JS. dihedral angles	& 0.205 & 0.244 \\
    \bottomrule
    \end{tabular}
    \label{tab: with_h}
\end{table}

\subsection{Comparison between \model{} and EDM on Molecules with All Elements}

Previously, we excluded some minor element types (B, Br, I, Si, Bi) from the GEOM-Drug dataset during data processing because they are not commonly used in drug design compared to other elements (C, N, O, F, P, S, Cl). Additionally, the generation of molecules containing these minor elements may bring more difficulties in subsequent biochemical experiments. Moreover, these minor elements accounted for less than 0.2\% of the GEOM-drug dataset, making it difficult for the models to properly learn their features. For instance, there were only 12 Si atoms in the entire GEOM-Drug dataset, and it was challenging for a model to learn how to generate Si atoms from these samples.

However, it is interesting to see how the model performs on the complete dataset. Therefore, we added an experiment to train our model on the complete GEOM-Drug dataset (i.e., all elements were used and hydrogens were not removed). The comparison of MolDiff and EDM on this complete dataset was shown in Tab. \ref{tab: all_ele}.

The performances of both \model{} and EDM decreased compared to those in the original filtered dataset (i.e., the dataset used in the main text). But \model{} still had high generation ability with a success rate of 0.860. However, the EDM model had a much lower success rate of 0.014, indicating that 100 samples resulted in about one valid and complete molecule, which was of little use for practical applications. Therefore, EDM cannot work on this complete GEOM-Drug dataset while \model{} still worked well. Regarding the qualities of generated molecules, the superiority of \model{} over EDM in the complete was the same as that in the filtered dataset. Therefore, working on the complete dataset, \model{} performed slightly worse than in the filtered dataset but still much better than EDM.

\begin{table}[htb]
    \centering
    \caption{Performance comparison on GEOM-Drug molecules with all elements}
    \begin{tabular}{c|cc}
    \toprule
        Model & MolDiff & EDM \\ \hline
        Validity & 0.947 & 0.029  \\
        Connectivity & 0.908 & 0.484  \\
        Succ. Rate & 0.860 & 0.014  \\
        Novelty & 1.000 & 1.000  \\
        Uniquness & 1.000 & 1.000  \\
        Diversity & 0.422 & 0.455  \\
        Sim. Val. & 0.696 & 0.668  \\\hline
        QED & 0.700 & 0.536  \\
        SA & 0.805 & 0.624  \\
        Lipinski & 4.874 & 4.839  \\ \hline
        RMSD. & 0.963 & 1.103  \\
        JS. bond lengths & 0.472 & 0.439 \\
        JS. bond angles & 0.178 & 0.435 \\
        JS. dihedral angles & 0.228 & 0.605 \\
    \bottomrule
    \end{tabular}
    \label{tab: all_ele}
\end{table}

\subsection{Comparison on QM9 dataset} \label{sec: comp_qm9}

QM9 dataset was widely adopted to benchmark previous 3D molecule generation models. We did not adopt it as our main dataset because the molecules are simpler and less drug-like than those in the GEOM-Drug dataset. Here we add a comparison on the QM9 dataset with more baselines, including EDM\cite{equivariant_diffusion}, EN-flow\cite{satorras2021enf}, G-Schnet\cite{gebauer2019symmetry}, and G-SphereNet\cite{luo2022autoregressive}. We followed the setup of EDM to split the QM9 dataset with train/validation/test size as 100k/18k/13k.

We compared the generation abilities and the accuracy of 3D geometries of these methods, as shown in Tab. \ref{tab: exp_qm9}. \model{} showed the best generation ability (with the highest validity, connectivity, and success rate), indicating that \model{} generated more valid and complete molecules than the others. Although EDM showed better performance in the QM9 dataset than it in the GEOM-Drug dataset, it still showed a relatively weaker generation ability than \model{}. Regarding the novelty, diversity, uniqueness, and similarity with the validation set, these methods achieved similar performance. For 3D geometries, \model{} achieved the best performance in terms of RMSD and the JS. divergence of dihedral angles and was second-best for the JS. divergence of bond lengths and bond angles, with the differences among methods being less significant. Overall, \model{} maintained its superior generation abilities and accuracy of 3D geometries on the QM9 dataset, although the margin over other methods was smaller due to the relative simplicity of molecules in QM9 compared to those in the GEOM-Drug dataset.

\begin{table}[htb]
    \centering
    \caption{Comparison on QM9 dataset}
    \begin{tabular}{c|ccccc}
    \toprule
         & MolDiff & EDM & EN-flow & G-SchNet & G-SphereNet  \\ \hline
        Validity & 0.970 & 0.925 & 0.407 & 0.876 & 0.161  \\
        Connectivity & 0.998 & 0.992 & 0.604 & 0.997 & 0.998  \\
        Succ. rate & 0.968 & 0.918 & 0.246 & 0.873 & 0.161  \\
        Novelty & 1.000 & 1.000 & 1.000 & 1.000 & 1.000  \\
        Diversity & 0.894 & 0.893 & 0.892 & 0.875 & 0.658  \\
        Uniqueness & 0.991 & 0.999 & 0.999 & 0.995 & 0.225  \\
        Sim. Val. & 0.705 & 0.602 & 0.490 & 0.628 & 0.732  \\ \hline
        RMSD & 0.193 & 0.249 & 0.587 & 0.278 & 0.199  \\
        JS. bond lengths & 0.340 & 0.231 & 0.513 & 0.342 & 0.367  \\
        JS. bond angles & 0.158 & 0.105 & 0.367 & 0.193 & 0.334  \\
        JS. dihedral angles & 0.368 & 0.383 & 0.713 & 0.492 & 0.721 \\
    \bottomrule
    \end{tabular}
    \label{tab: exp_qm9}
\end{table}


\end{document}